\newcommand{\pmu}{\partial_{\mu}}

\newcommand{\nmu}{\nabla_{\mu}}
\newcommand{\nmo}{\nabla^{\mu}}
\newcommand{\nnu}{\nabla_{\nu}}

\newcommand{\vpas}{4\pi\alpha_{{\rm S}}}

\newcommand{\tr}{{\rm tr}}

\newcommand{\MAmu}{{\mathcal A}_{\mu}}

\newcommand{\MAnu}{{\mathcal A}_{\nu}}
\newcommand{\MAlu}{{\mathcal A}_{\lambda}}

\newcommand{\exMAmu}{{}^{(ex)}\!{\mathcal A}_{\mu}}
\newcommand{\SMAmu}{{}^{(S)}\!{\mathcal A}_{\mu}}

\newcommand{\exAmu}{{}^{(ex)}\!A_{\mu}}

\newcommand{\Amu}{A_{\mu}}

\newcommand{\Aamu}{A^{a}_{\;\;\mu}}
\newcommand{\Aemu}{A^{1}_{\;\;\mu}}
\newcommand{\Aenu}{A^{1}_{\;\;\nu}}
\newcommand{\Azmu}{A^{2}_{\;\;\mu}}
\newcommand{\Aznu}{A^{2}_{\;\;\nu}}
\newcommand{\Bmu}{B_{\mu}}
\newcommand{\Bsmu}{B^{*}_{\mu}}
\newcommand{\Bmo}{B^{\mu}}
\newcommand{\Bsmo}{B^{*\mu}}
\newcommand{\Bnu}{B_{\nu}}

\newcommand{\Bsnu}{B_{\nu}^{*}}

\newcommand{\MDmu}{{\mathcal D}_{\mu}}
\newcommand{\MDnu}{{\mathcal D}_{\nu}}
\newcommand{\MDmo}{{\mathcal D}^{\mu}}

\newcommand{\MDlu}{{\mathcal D}_{\lambda}}

\newcommand{\Dmu}{D_{\mu}}
\newcommand{\Dmo}{D^{\mu}}

\newcommand{\eu}{{\rm e}^u}

\newcommand{\MFmunu}{{\mathcal F}_{\mu\nu}}

\newcommand{\MFnulu}{{\mathcal F}_{\nu\lambda}}
\newcommand{\MFlumu}{{\mathcal F}_{\lambda\mu}}

\newcommand{\SMFmunu}{{}^{(S)}\!{\mathcal F}_{\mu \nu}}

\newcommand{\exFmunu}{{}^{(ex)}\!F_{\mu \nu}}

\newcommand{\Fmunu}{F_{\mu \nu}}

\newcommand{\Famunu}{F^{a}_{\;\;\mu\nu}}
\newcommand{\Faomunu}{F^{\alpha}_{\;\;\mu\nu}}
\newcommand{\Faomulu}{F^{\alpha}_{\;\;\mu\lambda}}

\newcommand{\Faosulu}{F^{\alpha}_{\;\;\sigma\lambda}}

\newcommand{\Femunu}{F^{1}_{\;\;\mu\nu}}
\newcommand{\Fzmunu}{F^{2}_{\;\;\mu\nu}}
\newcommand{\Fdmunu}{F^{3}_{\;\;\mu\nu}}

\newcommand{\xefnu}{{}^{(xe)}\!f_{\nu}}
\newcommand{\Sfnu}{{}^{(S)}\!f_{\nu}}

\newcommand{\Gmunu}{G_{\mu\nu}}
\newcommand{\Gmonu}{G^{\mu}_{\;\;\nu}}

\newcommand{\Gsmunu}{G^{*}_{\mu\nu}}

\newcommand{\MHmu}{{\mathcal H}_{\mu}}

\newcommand{\MHmo}{{\mathcal H}^{\mu}}
\newcommand{\MHnu}{{\mathcal H}_{\nu}}
\newcommand{\MbHmu}{\bar{{\mathcal H}}_{\mu}}
\newcommand{\MbHmo}{\bar{{\mathcal H}}^{\mu}}
\newcommand{\MbHnu}{{\bar{\mathcal H}}_{\nu}}
\newcommand{\hmu}{h_{\mu}}

\newcommand{\hsmu}{h^{*}_{\mu}}

\newcommand{\hmo}{h^{\mu}}

\newcommand{\MI}{{\mathcal I}}

\newcommand{\SMJmu}{{}^{(S)}\!{\mathcal J}_{\mu}}

\newcommand{\jmu}{j_{\mu}}

\newcommand{\jnu}{j_{\nu}}
\newcommand{\jalomu}{j^{\alpha}_{\;\;\mu}}
\newcommand{\jaomu}{j^{a}_{\;\;\mu}}
\newcommand{\jaumu}{j_{a\mu}}
\newcommand{\jeomu}{j^{1}_{\;\;\mu}}

\newcommand{\jzomu}{j^{2}_{\;\;\mu}}

\newcommand{\jdomu}{j^{3}_{\;\;\mu}}
\newcommand{\jvomu}{j^{4}_{\;\;\mu}}
\newcommand{\jfomu}{j^{5}_{\;\;\mu}}
\newcommand{\jeonu}{j^{1}_{\;\;\nu}}
\newcommand{\jzonu}{j^{2}_{\;\;\nu}}
\newcommand{\jdonu}{j^{3}_{\;\;\nu}}

\newcommand{\jalumu}{j_{\alpha\mu}}

\newcommand{\jeumu}{j_{1\mu}}
\newcommand{\jzumu}{j_{2\mu}}
\newcommand{\jdumu}{j_{3\mu}}
\newcommand{\jvumu}{j_{4\mu}}
\newcommand{\jfumu}{j_{5\mu}}
\newcommand{\jbumu}{j_{\beta\mu}}
\newcommand{\jalonu}{j^{\alpha}_{\;\;\nu}}

\newcommand{\jbomo}{j^{\beta\mu}}
\newcommand{\exjmu}{{}^{(ex)}\!j_{\mu}}

\newcommand{\Kaubu}{K_{\alpha\beta}}

\newcommand{\Kaobo}{K^{\alpha\beta}}
\newcommand{\Kbogo}{K^{\beta\gamma}}
\newcommand{\keumu}{k_{1\mu}}
\newcommand{\kzumu}{k_{2\mu}}
\newcommand{\kdumu}{k_{3\mu}}

\newcommand{\Tmunu}{T_{\mu\nu}}
\newcommand{\MTmunu}{\mathcal T_{\mu\nu}}

\newcommand{\esTmunu}{{}^{(es)}\!T_{\mu\nu}}
\newcommand{\TTmunu}{{}^{(T)}\!T_{\mu\nu}}
\newcommand{\DTmunu}{{}^{(D)}\!T_{\mu\nu}}
\newcommand{\GTmunu}{{}^{(G)}\!T_{\mu\nu}}

\newcommand{\Mvaumu}{\upsilon_{\alpha\mu}}

\newcommand{\Gmu}{\Gamma_{\mu}}

\newcommand{\Gnu}{\Gamma_{\nu}}

\newcommand{\talu}{\tau_{\alpha}}
\newcommand{\tbu}{\tau_{\beta}}

\newcommand{\D}{{\mathcal{D}}}

\newcommand{\I}{{\mathcal{I}}}

\newcommand{\vecr}{\vec{r}}

\newcommand{\exAo}{{}^{(ex)}\!A_0}
\newcommand{\eAo}{{}^{(1)}\!A_0}
\newcommand{\zAo}{{}^{(2)}\!A_0}
\newcommand{\dAo}{{}^{(3)}\!A_0}

\newcommand{\aAo}{{}^{(a)}\! A_0}

\newcommand{\alo}{{}^{(a)}\!l_0}

\newcommand{\kaumu}{k_{a\mu}}
\newcommand{\eko}{{}^{(1)}\! k_0}
\newcommand{\zko}{{}^{(2)}\! k_0}
\newcommand{\dko}{{}^{(3)}\! k_0}
\newcommand{\ako}{{}^{(a)}\! k_0}

\newcommand{\nlu}{\nabla_{\!\lambda}}

\newcommand{\MBmu}{\mathcal{B}_{\mu}}

\newcommand{\Bkomu}{B^k_{\;\;\mu}}
\newcommand{\Bskomu}{B^{\ast k}_{\;\;\;\;\mu}}

\newcommand{\MFmono}{{\mathcal F}^{\mu\nu}}

\newcommand{\MbTmunu}{\bar{{\mathcal T}}_{\mu\nu}}
\newcommand{\emMAmu}{{}^{(em)}\!\mathcal{A}_{\mu}}
\newcommand{\Admu}{A^{3}_{\;\;\mu}}
\newcommand{\Adnu}{A^{3}_{\;\;\nu}}
\newcommand{\DGmu}{\textnormal I \! \Gamma_\mu}
\newcommand{\DGmo}{\textnormal I \! \Gamma^\mu}
\newcommand{\DGnu}{\textnormal I \! \Gamma_\nu}
\newcommand{\DGlu}{\textnormal I \! \Gamma_\lambda}

\documentclass[amssymb,amsmath,preprint,eqsecnum,aps,nofootinbib,showpacs]{revtex4}
\usepackage{epsfig,amssymb,amsmath,stmaryrd,dsfont,color}
\usepackage[a4paper]{geometry}
\newcommand{\bdot}{\, {\scriptscriptstyle{}^{\bullet}} \,}

\begin{document}

\title{\LARGE{Positive and Negative Charges\\ in \\ Relativistic Schr\"odinger Theory}\\[6ex]}
\author{T. Beck and M. Sorg}
\affiliation{ II.\ Institut f\"ur Theoretische Physik der
Universit\"at Stuttgart\\Pfaffenwaldring 57\\ D-70550 Stuttgart\\ Germany\\
{\rm e-mail:} {\tt sorg@theo2.physik.uni-stuttgart.de}}

\begin{abstract}
\indent Relativistic Schr\"odinger Theory (RST), as a general gauge theory for the description of relativistic N-particle systems, is shown to be a mathematically consistent and physically reasonable framework for an arbitrary assemblage of positive and negative charges. The electromagnetic plus exchange interactions within the subset of {\it identical} particles are accounted for in a consistent way, whereas {\it different} particles can undergo only the electromagnetic interactions. The origin of this different interaction mechanism for the subsets of identical and non-identical particles is traced back to the fundamental conservation laws for charge and energy-momentum: in order that these conservation laws can hold also for different particles, the structure group $\mathcal U(N)$ of the fibre bundles must be reduced to its maximal Abelian subgroup $\mathcal U(1) \times \mathcal U(1) \times \dots \times \mathcal U(1)$, which eliminates the exchange part of the bundle connection. The persisting Abelian gauge symmetry adopts the meaning of the proper gauge group for the electromagnetic interactions which apply to the identical and non-identical particles in the same way. Thus in RST there is an intrinsic dynamical foundation of the emergence of exchange effects for identical particles, whereas the conventional theory is invaded by the exchange phenomenon via a purely kinematical postulate, namely the antisymmetrization postulate for the wave functions due to Pauli's exclusion principle. As a concrete demonstration, a three-particle system is considered which consists of a positively charged particle of arbitrary rest mass and of two negatively charged particles of equal spin, mass and charge (e.g. electrons).\vspace{3cm}
\end{abstract}

\maketitle

\section{Introduction and Survey of Results}

\indent There are many instances where the behaviour of quantum particles does appear very curious to the classical observer; but surely the strangest phenomenon refers to the effect of particle entanglement. Indeed this phenomenon implies the existence of very unusual correlations and exchange effects which are far beyond the scope of classical physics. On the other hand, the observational implications of this exotic behaviour of quantum matter are generally accepted so that any form of quantum theory is obliged to take account of these effects. To this end, the standard theory resorts to the Pauli-exclusion principle for fermions and recasts this into mathematical terms by postulating the antisymmetrization of the many-particle wave functions (or postulating anti-commutation relations for the fermionic field operators, resp.), see ref.\cite{b22} for a historical survey. However these endeavours did not result in a powerful and generally accepted quantum mechanics for the relativistic $N$-particle systems (e.g. many-electron atoms), see the many critical remarks about the Bethe-Salpeter equations \cite{b31,a23} as a possible candidate. Consequently one has to resort to approximative methods such as the relativistic $1/Z$-expansion method \cite{a6}, relativistic many-body perturbation theory (MBPT) \cite{b15,a4}, all-order technique in MBPT \cite{a5}, or the multi-configuration Dirac-Fock method (MCDF) \cite{b13,b14,a3}.

\indent A fresh approach to this old problem of setting up a consistent relativistic quantum mechanics (not quantum field theory!) for the N-particle systems with inclusion of the exchange effects has recently been put forward in form of the Relativistic Schr\"odinger Theory (RST) \cite{a43,u5,u6}. This approach is a relativistic gauge field theory of fluid-dynamic character (such as the well-known density functional theory) and is based upon the structure group $\mathcal U(N)$ which is spontanously broken down to its maximal Abelian subgroup $\mathcal U(1) \times \mathcal U(1) \dots \mathcal U(1)$ working then as the proper gauge group for the {\it electromagnetic} interactions. The frozen gauge degrees of freedom refer to the {\it exchange} interactions which thus are treated as real forces on the same footing as the electromagnetic interactions. Just as the electromagnetic field is generated by the electromagnetic currents of the particles, the exchange field is generated by the exchange currents which are different from zero only in those regions of space-time where the individual particle wave-functions do overlap. This is a {\it dynamical} mechanism of particle entanglement \cite{a43}, whereas the conventional mechanism is rather of {\it kinematical} origin, namely through the antisymmetrical superposition of products of wave functions.

\indent Thus, RST appears to be well suited to take account of the notorious exchange effects among identical particles; but this is not enough for a complete theory of elementary matter. Rather, such a theory should also be able to explain the reason why {\it non-identical} particles do {\it not} undergo the exchange interactions. In the conventional theory, this problem is solved again in a purely kinematical way: the non-identical particles are simply not included in the antisymmetrization procedure. However in RST, the non-existence of exchange effects between particles of different mass and charge is tied up more intimately to the physics (i.e conservation laws) of the N-particle systems. More concretely, in order that the continuity equations for charge and energy-momentum be valid for a mixed assembly of both identical and different particles, the strucure group must be reduced to some smaller subgroup. This implies that the bundle connection $\mathcal A_\mu$ ("gauge potential") takes its values in a subalgebra of the original structure algebra $\mathfrak u(N)$ for identical particles. Here, it turns out that the redundand connection components, which are to be dropped for the reduction process, are just those exchange potentials which would have mediated the exchange interactions between the non-identical particles. Thus, in RST there exists a natural intrinsic mechanism for suppressing the unwanted exchange interactions, whereas in the conventional theory this type of force is avoided simply by not including the different particles' wave functions into the antisymmetrization postulate.

\indent The intention of the present paper is to present a detailed elaboration of the specific way in which the exchange interactions emerge in RST for the identical particles but are suppressed for the non-identical particles. The corresponding results are worked out in the following arrangement:

\indent In {\bf Sect.II} the fundamentals of RST are presented for a three-particle system consisting of a positively charged particle (e.g. positron or proton) and two negatively charged particles (e.g. electrons). The point of departure is here the {\it total velocity operator} $\DGmu$ (\ref{2.8}) which is chosen as the charge-weighted direct sum of the standard Dirac matrices $\gamma_\mu$. The central meaning of the velocity operator becomes evident from the fact that it implies the emergence of the {\it total current density} $j_\mu$ (\ref{2.10})-(\ref{2.11}) and of the energy-momentum density $\DTmunu$ (\ref{2.14})-(\ref{2.15}) which both are required to obey certain conservation laws, e.g. the continuity equation in the local formulation (\ref{2.3}). The point with these conservation laws is now that they require the covariant constancy of both the total velocity operator $\DGmu$ (\ref{2.42}) and the mass operator $\mathcal M$ (\ref{2.46}). However, in order to ensure this required covariant constancy of the objects $\DGmu$ and $\mathcal M$, the structure group $\mathcal U(3)$ for three identical particles must be reduced to $\mathcal U(1) \times \mathcal U(2)$ for a system of two identical particles (i.e. the electrons) and one different particle (i.e. positron or proton).

\indent This reduction process with all its implications is described in detail in {\bf Sect. III}. The associated reduced connection $\SMAmu$ is shown by equations (\ref{3.3})-(\ref{3.4}) where the exchange potentials referring to the different particles become cut out so that the remaining exchange potential $\mathcal B_\mu$ exclusively refers to the two-identical particles (i.e. the electrons). Clearly such a reduction process has its consequences also for the gauge field dynamics (see eqns. (\ref{3.9a})-(\ref{3.9d})), the fibre metric $K_{\alpha\beta}$ (\ref{3.23}) in the associated Lie algebra bundle as the coupling matrix of the gauge field modes to the matter currents, and for the matter dynamics itself, see the coupled Dirac system (\ref{3.32a})-(\ref{3.32c}) with the gauge covariant derivatives of the matter fields $\psi_a$ being presented through equations (\ref{3.31a})-(\ref{3.31c}).

\indent Whether or not the present construction of the exchange interactions has experimental relevance, may be tested by inspecting the energy spectrum of the stationary bound states. Therefore the general RST energy eigenvalue problem is set up and found to be constituted essentially by three subproblems: (i) {\it mass eigenvalue equations} for the wave functions $\psi_a(\vec r)$ (see equations (\ref{4.2})-(\ref{4.4b}), (ii) {\it non-Abelian Poisson equations} for the gauge fields ($\rightarrow$ equations (\ref{4.72a})-(\ref{4.73d})), (iii) and the {\it total energy functional} $E_T$ (\ref{4.37}) whose value upon the coupled matter and gauge field configuration yields the desired energy eigenvalues of the stationary bound systems. However, here does arise now a very critical point for the RST framework which must be successfully overcome in order that RST can be claimed to represent a generally valid theory for arbitrarily composed N-particle systems.

\indent The point here is that positive and negative charges do react in a different way to external potentials acting upon them. For instance, if the N-particle system as a whole is acted upon by an external potential $\exAmu$, the positively charged members of the system receive an oppositely directed force as compared to the negatively charged members. In this sense, the first test is successfully passed by the present construction, see the discussion below equations (\ref{4.49a})-(\ref{4.49c}).

\indent Next, choose a positively (negatively) charged member of the $N$-particle system and consider its electromagnetic interaction energy with any other positive or negative charge of the assembly. Clearly, when both selected particles carry a like (different) charge, then the interaction energy must be positive (negative). However these two-particle relations are also found to be correctly incorporated in the present RST framework, see the discussion below equation (\ref{4.17}).

\indent A further crucial point refers to the non-relativistic approximation of the RST energy eigenvalue problem. If the theory can describe the relativistic effects correctly, its non-relativistic limit must coincide with the well-established non-relativistic quantum mechanics (i.e. Schr\"odinger equation or Hartree-Fock equation, resp.). Also this requirement is found to be satisfied by the present construction, see the discussion below equation (\ref{4.29}) and in connection with the kinetic energies (\ref{4.32})-(\ref{4.34b}). In this context, there arises a certain peculiarity: the mass eigenvalue of the positively charged particle turns out to be negative, see equation (\ref{4.6}). However a careful inspection of the kinetic energy of the positive charge demonstrates that this quantity is always positive and both the positive and negative charges carry always positive energy located in their matter fields $\psi_a(\vec r)$, see the discussion of equations (\ref{4.62a})-(\ref{4.62b}).

\indent Summarizing, one finds that RST can deal with an assembly of positive and negative charges in a way which meets with all physical requirements, and especially takes account for the exchange effects and satisfies also all conservation laws which are expected to hold for such systems. This may be considered as a necessary theoretical presumption for entering now the field of concrete numerical calculations in atomic and molecular physics.

\section{RST Fundamentals}

The general features of RST have been presented in detail already in some preceding papers \cite{a43, u5, u6} so that it may be sufficient to mention here only the most important facts, being necessary to make the subsequent discussions sufficiently self-contained. Thus, we may restrict ourselves to those key points which have to emerge in any field theory of elementary particles: conservation laws, kinematics and dynamics, and action principle.

\subsection{Conservation Laws}

Concerning the general motion of matter, the most striking fact refers to the existence of certain conservation laws. Here, one of the most important ones is the {\it charge conservation} of a system of N particles. In order to put this into mathematical terms, one introduces the {\it total current} $\jmu$ and subjects this object to the local conservation law
\begin{equation}\label{2.1}
\nabla^\mu \, \jmu = 0 \ ,
\end{equation}
so that the total charge $z$ as an integral over some hypersurface (S) becomes independent of just that hypersurface
\begin{equation}\label{2.2}
z = \int_{(S)} \jmu \, dS^\mu \ .
\end{equation}
\indent A further conservation law of a {\it closed} N-particle system refers to its energy-momentum content. The corresponding global law, such as for charge conservation (\ref{2.2}), can again be ensured by demanding locally for the {\it total energy-momentum density} $\TTmunu$
\begin{equation}\label{2.3}
\nabla^\mu \, \TTmunu = 0 \ .
\end{equation}
However, when the N-particle system is not closed but is acted upon by some external electromagnetic field strength $\exFmunu$ excited by an external current ${}^{(ex)} \! f_\nu$, the source of $\TTmunu$ is not zero but equals the Lorentz force density ${}^{(xe)} \! f_\mu$
\begin{equation}\label{2.4}
\nabla^\mu \, \TTmunu = - {}^{(xe)} \! f_\nu \ .
\end{equation}
In general, the force density ${}^{(xe)} \! f_\nu$ is composed of the external current ${}^{(ex)} \! j_\mu$ and an internal field strength $F_{\mu\nu}$ (see below) and is of the expected Lorentzian form
\begin{equation}\label{2.5}
\xefnu = -\hbar c \, F_{\mu\nu} \, {}^{(ex)}\!j^\mu \ .
\end{equation}

\subsection{RST Kinematics}

The second step for the construction of RST consists in introducing the fundamental (but unobservable) fields of the theory which build up the observable densities (such as $j_\mu$ or $\Tmunu$). The fundamental matter field $\Psi(x)$ of RST is chosen to be a (local) section of some complex vector bundle over space-time as the base space. In the case of N Dirac particles, the typical vector fibre of such a bundle is a (4N)-dimensional representation space of the Clifford algebra $\mathbb{C}(1,3)$
\begin{equation}\label{2.7}
\DGmu \DGnu + \DGnu \DGmu = 2 \, g_{\mu\nu} \cdot \mathds{1}
\end{equation}
where the {\it total velocity operators} $\DGmu$ are the direct sum of the ordinary (i.e four-dimensional) Dirac matrices $\gamma_\mu$
\begin{equation}\label{2.8}
\DGmu = (-\gamma_\mu) \oplus \gamma_\mu \oplus \gamma_\mu \dots \ .
\end{equation}
The distributions of signs in the direct sum (\ref{2.8}) reflects the distribution of positive and negative charges over the considered N-particle system. Thus, the explicit arrangement (\ref{2.8}) describes a three-particle system where the first particle is taken (by convention) to be positively charged (e.g. positron or proton) and the second and third particles carry the same negative charge (e.g. two electrons). In the preceding papers \cite{a43, u5, u6} we considered systems with N {\it identical} particles (i.e. electrons) which have a total velocity operator $\Gamma_\mu$ of the following form
\begin{equation}\label{2.9}
\Gamma_\mu = \gamma_\mu \oplus \gamma_\mu \oplus \gamma_\mu \dots \ .
\end{equation}

By this construction, the total current $j_\mu$ can be defined now in terms of a {\it pure state} $\Psi$ and of the velocity operator $\DGmu$ (\ref{2.8}) as
\begin{equation}\label{2.10}
j_\mu = \bar\Psi \, \DGmu \, \Psi \ ,
\end{equation}
or in case of a mixture
\begin{equation}\label{2.11}
j_\mu = \tr \left(\mathcal{I} \cdot \DGmu \right) \ .
\end{equation}
Here the {\it intensity matrix} $\mathcal I$ may be parametrized by N pure states $\Psi_n$ ($n=1 \dots N$) as
\begin{equation}\label{2.12}
\mathcal{I} = \Psi_1 \otimes \bar\Psi_1 + \Psi_2 \otimes \bar\Psi_2 + \dots \, \Psi_N \otimes \bar\Psi_N \ .
\end{equation}
In this way, the N-particle bundle arises as the Whitney sum of the individual one-particle bundles. This is in contrast to conventional quantum theory which is of statistical character and therefore uses the tensor product of Hilbert spaces; whereas RST is a fluid-dynamic approach (similar to the density functional theory \cite{b57,b58}) where the sum construction is more adequate.

\indent Once an observable density such as the total current $j_\mu$ (\ref{2.10})-(\ref{2.11}) has been defined, one will apply this recipe of constructing the physical densities also to the energy-momentum density of the Dirac matter ($\DTmunu$, say). Accordingly, one introduces an {\it energy-momentum operator} $\MTmunu$ such that the matter density $\DTmunu$ appears as
\begin{equation}\label{2.13}
\DTmunu = \tr \left( \mathcal{I} \cdot \mathcal \Tmunu \right) \ ,
\end{equation}
or, resp., for the pure states
\begin{equation}\label{2.14}
\DTmunu = \bar \Psi \, \mathcal \Tmunu \, \Psi \ .
\end{equation}
Clearly there arises now the question of relationship between the total velocity operator $\DGmu$ (\ref{2.8}) and the energy-momentum operator $\mathcal \Tmunu$ (\ref{2.13}). In order to establish the link between these two objects, one introduces first the {\it Hamiltonian} $\mathcal{H}_\mu$, a $gl(N, \mathbb{C})$-valued one-form, which admits to express the (pseudo-Hermitian) energy-momentum operator $\mathcal \Tmunu$ ($ = \MbTmunu$) as
\begin{equation}\label{2.15}
\MTmunu = \frac{1}{4} \, \big\{ \DGmu \MHnu + \MbHnu \DGmu + \DGnu \MHmu + \MbHmu \DGnu \big\} \ .
\end{equation}

\indent Most modern field theories are gauge theories and this is true also for RST. The {\it structure group} (i.e. formal gauge group) for an N-particle system is (a 4N-dimensional representation of) the  unitary group $\mathcal U(N)$, or a subgroup thereof, which however is spontaneously broken down to its maximal Abelian subgroup $\mathcal{U}(1) \times \mathcal{U}(1) \times \dots \times \mathcal{U}(1)$. Thus, the whole set of $N^2$ generators $\tau_\alpha$ ($\alpha = 1 \dots N^2$) is subdivided into the (anti-Hermitian) {\it electromagnetic generators} $\tau_a = -\bar{\tau}_a$ ($a=1 \dots N$) and the complementary {\it exchange generators} $\chi_k, \bar{\chi}_k$ ($k=1 \dots N(N-1)/2$). The residual gauge symmetry $\mathcal{U}(1) \times \mathcal{U}(1) \times \dots \times \mathcal{U}(1)$ refers to the electromagnetic interactions among the particles, whereas the frozen gauge degrees of freedom describe their exchange interactions. Accordingly, the $\mathfrak{u}(N)$-valued bundle-connection ({\it "gauge potential"}) $\MAmu$ is split up into three parts: external potential $\exMAmu$, electromagnetic potential ${}^{(em)}\!{\mathcal A}_{\mu}$ and exchange potential $\MBmu$:

\begin{equation}\label{2.16}
\MAmu = \exMAmu + {}^{(em)}\!{\mathcal A}_{\mu} + \MBmu
\end{equation}
with the individual constitutents being given by
\begin{subequations}\label{2.17}
\begin{align}
\exMAmu & = - i \exAmu \cdot \mathds 1 \label{2.17a}\\
\emMAmu & = \Aamu \tau_a, \ (a=1 \dots N) \label{2.17b}\\
  \MBmu & = \Bkomu \chi_k - \Bskomu \bar{\chi}_k \ , \ \big(k=1\dots N(N-1)/2\big) \ . \label{2.17c}
\end{align}
\end{subequations}
The bundle connection $\MAmu$ is adopted to be anti-Hermitian ($\bar{{\MAmu}} = - \MAmu$) and its electromagnetic part ${}^{(em)}\mathcal{A}_{\mu} $ is assumed to commute with the total velocity operator $\DGmu$, i.e. we require
\begin{equation}\label{2.18}
\left[ \tau_a , \DGmu \right] =0 \ .
\end{equation}

A similar splitting as for the connection $\MAmu$ does then apply also to the bundle curvature $\MFmunu$ ({\it field strength}), i.e. we put
\begin{subequations}\label{2.19}
\begin{align}
\MFmunu & =      -i\ \exFmunu \cdot \mathds{1} + \Famunu \tau_a + \Gmunu \chi -
                     \Gsmunu\bar{\chi} \label{2.19a} \\
				& \equiv -i\ \exFmunu \cdot \mathds{1} + \Faomunu \tau_{\alpha} \ . \label{2.19b}
\end{align}
\end{subequations}
The curvature $\MFmunu$ emerges in those {\it bundle identities} like
\begin{subequations}\label{2.20}
\begin{align}
      \left[ \MDmu \MDnu - \MDnu \MDmu \right] \, \Psi & \equiv \MFmunu \, \Psi \label{2.20a}\\
\left[ \MDmu \MDnu - \MDnu \MDmu \right] \, \mathcal I & \equiv \left[ \MFmunu , \mathcal I \right] \ , \label{2.20b}
\end{align}
\end{subequations}
or also in the well-known {\it Bianchi identity}
\begin{eqnarray}\label{2.21}
\MDlu\MFmunu &+& \MDmu\MFnulu + \MDnu\MFlumu \equiv 0 \\
\big(\MDlu\MFmunu\Big. &\doteqdot& \Big.\nlu\MFmunu + \left[\MAlu,\MFmunu\right]\big) \ . \nonumber
\end{eqnarray}
The latter identity is implied by the fact that the curvature $\MFmunu$ is the (non-Abelian) curl of the connection $\MAmu$:
\begin{equation}\label{2.22}
\MFmunu \doteqdot \nmu\MAnu - \nnu\MAmu + \left[\MAmu,\MAnu \right] \ .
\end{equation}
An important identity of the bundle geometry is also the following:
\begin{equation}\label{2.23}
\left[\MDmu,\MDnu \right]\, \MFmono \equiv 0 \ .
\end{equation}
Indeed, as we will readily see, this is a severe restriction to the gauge field dynamics which has to specify some field equation for the connection $\MAmu$.

\indent The preceding gauge structure of RST gives also rise to the introduction of the {\it gauge velocity operators} $\Mvaumu$ which are defined in terms of the $\mathcal U(N)$ generators $\tau_\alpha$ and total velocity operator $\DGmu$ as follows
\begin{equation}\label{2.24}
\Mvaumu = \frac{i}{2} \left\{ \talu, \DGmu \right\} \ .
\end{equation}
Clearly, these objects induce the emergence of the {\it RST currents} $\jalumu$ in a way quite analogous to the total current $\jmu$ (\ref{2.10}):
\begin{equation}\label{2.25}
\jalumu \doteqdot \tr \left( \I \cdot \Mvaumu \right) \ ,
\end{equation}
or, resp., for the case of a pure state $\Psi$
\begin{equation}\label{2.26}
\jalumu \doteqdot \bar \Psi \, \Mvaumu \, \Psi \ .
\end{equation}
One cannot expect these RST currents $\jalumu$ to obey a conservation law such as the total current $j_\mu$ (\ref{2.1}), but one may well suppose that they might obey a continuity equation of the kind
\begin{equation}\label{2.27}
D^\mu \jalumu \equiv 0 \ .
\end{equation}
Since the RST currents $\jalumu$ formally transform under the adjoint representation of the structure group $\mathcal U(N)$, the covariant derivative ($D$) in (\ref{2.27}) is to be defined as follows:
\begin{equation}\label{2.28}
D_\lambda \jalumu = \nabla_\lambda \jalumu - \omega^\beta_{\ \alpha\mu} \, \jbumu \ .
\end{equation}
Here the connection one-form $\omega_\mu = \left\{ \omega^\beta_{\ \alpha\mu} \right\}$ constitutes the adjoint representation of the "internal" connection $\SMAmu$ ($\doteqdot \emMAmu + \MBmu$), i.e.
\begin{equation}\label{2.29}
\omega^\beta_{\ \alpha\mu} = C^\beta_{\ \gamma\alpha} \, A^\gamma_{\ \mu} \ ,
\end{equation}
where the structure constants $C^\beta_{\ \gamma\alpha}$ are defined as usual
\begin{equation}\label{2.30}
\left[ \tau_\gamma , \tau_\alpha \right] = C^\beta_{\ \gamma\alpha} \, \tau_\beta \ .
\end{equation}
\indent Finally, before the RST dynamics can be established in such a way that all the present conservation laws and balance equations (\ref{2.1}), (\ref{2.3}), (\ref{2.4}) and (\ref{2.27}) become automatically satisfied, one has to consider a further kinematical requirement, i.e. the Lorentz invariance. This symmetry is (besides the gauge structure) an indispensible feature of all the successful modern particle theories. However it is easy to see that the present RST kinematics is actually Lorentz invariant. In order to give a simple demonstration, return for a moment to the total current $j_\mu$ (\ref{2.10}) and recall that the Dirac spinor $\Psi$ transforms under the group $Spin(1,3)$ whenever the space-time coordinates are subject to a Lorentz transformation as an element of the pseudo-orthogonal group $SO(1,3)$. More concretely, define first the $Spin(1,3)$ generators $\Sigma_{\mu\nu}$ ($= -\bar{\Sigma}_{\mu\nu}$) in terms of the Dirac matrices $\Gmu$ (\ref{2.9}) through
\begin{equation}\label{2.31}
\Sigma_{\mu\nu} \doteqdot \frac{1}{4} \left[ \Gmu, \Gnu \right]
\end{equation}
so that the following commutation relations do hold:
\begin{equation}\label{2.32}
\left[ \Sigma_{\mu\lambda} , \DGnu \right] = g_{\lambda\nu} \DGmu -g_{\mu\nu} \DGlu \ .
\end{equation}
Thus, the homomorphic counterpart $\Lambda$ in the Lorentzgroup $SO(1,3)$ with generators $\mathcal L_{\mu\nu}$ ($= -\mathcal L^\dagger_{\mu\nu}$)
\begin{equation}\label{2.33}
\Lambda = \exp\left[\frac{1}{2} a^{\mu\nu} \mathcal L_{\mu\nu} \right] \ ,
\end{equation}
being due to the $Spin(1,3)$ element $\mathcal S$
\begin{equation}\label{2.34}
\mathcal S = \exp\left[\frac{1}{2} a^{\mu\nu} \, \Sigma_{\mu\nu} \right] \ ,
\end{equation}
obeys the adjoint representation law
\begin{equation}\label{2.35}
\bar{\mathcal S} \, \DGmu \, \mathcal S = ( \Lambda^{-1} )^\nu_{\ \mu} \, \DGnu \ .
\end{equation}
Consequently, a Lorentz transformation $\Lambda$ (\ref{2.33}) of the space-time coordinates induces the associated spinor transformation
\begin{equation}\label{2.36}
\Psi \Rightarrow \Psi' = \mathcal S \Psi
\end{equation}
which then recasts the total current $j_\mu$ (\ref{2.10}) into the following expected form:
\begin{equation}\label{2.37}
j_\mu \Rightarrow j_\mu' = \bar \Psi \left( \bar{\mathcal S } \, \DGmu \, \mathcal S \right) \Psi = ( \Lambda^{-1} )^\nu_{\ \mu} \, j_\nu \ .
\end{equation}
The correct transformation behaviour of all the other RST objects may be checked in an analogous way.

\indent Thus, the RST kinematics appears as a logically consistent theoretical framework, and one can go now to set up the RST dynamics in such a way that all these kinematical prerequisites are respected.

\subsection{RST Dynamics}
\indent According to the subdivision of the RST variables into matter fields ($\Psi$) and gauge fields ($\MAmu$), the dynamics must necessarily appear as a coupled set of matter and gauge field equations. In order that both kinds of field equations be consistent with each other, one needs a {\it "compatibility condition"} which will emerge in a most natural way in form of a fibre metric $\Kaubu$ for the associated Lie algebra bundle.
\indent First turning to the matter part of the dynamics, one chooses the \underline{R}elativistic von \underline{N}eumann \underline{E}quation (RNE) as the field equation for the intensity matrix $\mathcal I$
\begin{equation}\label{2.38}
\MDmu \, \MI = \frac{i}{\hbar c}\left[ \MI \, \bar{\MHmu} - \MHmu \, \MI \right] \ ,
\end{equation}
or, resp., for the situation of a pure state $\Psi$ one takes the \underline{R}elativistic \underline{S}chr\"odinger \underline{E}quation (RSE)
\begin{equation}\label{2.39}
i\hbar c \, \MDmu \Psi = \MHmu \Psi \ .
\end{equation}
Clearly, both matter equations (\ref{2.38}) and (\ref{2.39}) must be consistent when the intensity matrix $\mathcal I$ (\ref{2.12}) is decomposed into the tensor product of pure states $\Psi_n$ which themselves obey the RSE (\ref{2.39}).

\indent The Hamiltonian $\mathcal H_\mu$ itself is a dynamical object of the theory and therefore requires the specification of some field equations of its own. Here, the first constraint to be satisfied refers to the bundle identities (\ref{2.20a})-(\ref{2.20b}) and therefore the first field equation for $\MHmu$ is the {\it integrability condition}
\begin{equation}\label{2.40}
\MDmu \MHnu - \MDnu \MHmu + \frac{i}{\hbar c} \left[\MHmu,\MHnu \right] = i\hbar c \, \MFmunu \ .
\end{equation}
Next, one has to guarentee the validity of those continuity equations for the currents $j_\mu$ (\ref{2.1}) and $\jalumu$ (\ref{2.27}). Indeed, this can be achieved by requiring the commutativity of the total velocity operator $\DGmu$ and the structure algebra \cite{a43}
\begin{align}\label{2.41}
[ \tau_\alpha,& \, \DGmu  ] = 0 \\
(1 \leq \,& \alpha \leq N'^2) \ , \nonumber
\end{align}
which together with the condition of covariant constancy
\begin{equation}\label{2.42}
\MDmu \, \DGnu \equiv 0
\end{equation}
renders the total velocity operator $\DGmu$ as an absolute (i.e. non-dynamical) object of the theory, similarly to the Dirac matrices $\gamma_\mu$ in the conventional quantum theory. Furthermore it is necessary to introduce a (Hermitian) {\it mass operator} $\mathcal M$ ($=\bar{\mathcal M})$ through
\begin{equation}\label{2.43}
\mathcal M c^2 \doteqdot \MbHmo \, \DGmu = \DGmu \, \MHmo
\end{equation}
which must commute with both the reduced structure algebra $\mathfrak{u}(N')$ of dimension $N'\leq N$
\begin{align}\label{2.44}
\left[ \tau_\alpha, \mathcal M \right] = 0 \\
(\alpha = 1 \dots N'^2 ) \nonumber
\end{align}
and with the total velocity operator $\DGmu$
\begin{equation}\label{2.45}
\left[ \DGmu, \mathcal M \right] = 0 \ .
\end{equation}
Furthermore we have to demand also that the mass operator $\mathcal M$ be covariantly constant
\begin{equation}\label{2.46}
\MDmu \mathcal M \equiv 0
\end{equation}
so that, in view of the commutativity (\ref{2.44}), $\mathcal M$ counts also as an absolute object of the theory. Clearly, the particle masses of the theory must be fixed from the outside and cannot be determined from the dynamics itself (similarly to the particle charges).
In general, the conditions (\ref{2.44})-(\ref{2.45}) reduce the structure group $\mathcal U(N)$ of {\it identical} particles to some smaller subgroup of dimension $N' \leq N$, but these conditions are then sufficient in order to ensure the desired continuity equations for the currents $j_\mu$ and $\jalumu$.

\indent However, if one wants to have ensured also the energy-momentum conservation or, resp., balance equations (\ref{2.3})-(\ref{2.4}), one needs an additional constraint upon the Hamiltonian $\MHmu$, namely the following {\it conservation equation}
\begin{equation}\label{2.47}
\MDmo \MHmu - \frac{i}{\hbar c} \, \MHmo \MHmu = -i\hbar c \Big\{ \Big(\frac{\mathcal M c}{\hbar} \Big)^2 + \Sigma^{\mu\nu} \MFmunu \Big \} \ .
\end{equation}
This together with the integrability condition (\ref{2.40}) constitutes the {\it Hamiltonian dynamics}. This is an interesting relation in so far as it does not only ensure the validity of energy-momentum conservation, but it implies also the \underline{K}lein \underline{G}ordon \underline{E}quation (KGE) for the wave function $\Psi$
\begin{equation}\label{2.48}
\MDmo \MDmu \Psi + \Big(\frac{\mathcal M c}{\hbar} \Big)^2 \Psi = - \Sigma^{\mu\nu} \MFmunu \Psi \ .
\end{equation}
Indeed this equation can easily be deduced from the RSE (\ref{2.39}) by simply differentiating once more and eliminating the Hamiltonian $\MHmu$ just by use of the conservation equation (\ref{2.47}). Furthermore, one can show, that this conservation equation (\ref{2.47}) is equivalent to the former relation (\ref{2.43}) which itself may again be used in order to eliminate the Hamiltonian $\MHmu$ from the RSE (\ref{2.39}) which then results in the well-known \underline{D}irac \underline{E}quation (DE):
\begin{equation}\label{2.49}
i \hbar \, \DGmo \MDmu \Psi = \mathcal M c \, \Psi \ .
\end{equation}
This pleasant result says that Dirac's proposition for the wave equation of relativistic spin-$\frac{1}{2}$ particles emerges from the RST kinematics as a kind of minimal program for satisfying the fundamental conservation laws! ($\leadsto$ Dirac realization of RST).

\indent Once the Dirac equation has thus been established within the RST framework, it can be used in order to further eliminate the Hamiltonian $\MHmu$ for the pure states (but not for the mixtures). For instance, reconsider the matter energy-momentum density $\DTmunu$ being given by equation (\ref{2.14}). Now using again the DE (\ref{2.49}) for eliminating the Hamiltonian $\MHmu$ yields for the matter density $\DTmunu$:
\begin{equation}\label{2.50}
\DTmunu = \frac{i \hbar c}{4} \left\{ \bar \Psi \DGmu (\MDnu \Psi) - (\MDnu \bar \Psi) \DGmu \Psi + \bar \Psi \DGnu (\MDmu \Psi) - (\MDmu \bar \Psi) \DGnu \Psi \right \} \ .
\end{equation}
As a nice exercise, one may calculate here the divergence of $\DTmunu$ by extensive use of all the preceding RST kinematics and dynamics (especially KGE (\ref{2.48}) and DE (\ref{2.49}) ) in order to verify the following energy-momentum balance:
\begin{equation}\label{2.51}
\nabla^\mu {}^{(D)}\!T_{\mu\nu} = \hbar c \, \big\{ {}^{(ex)}\!F_{\mu\nu} j^\mu + F^\alpha_{\ \mu\nu} j_\alpha^{\ \mu} \big\} \ ,
\end{equation}
with the curvature components $F^\alpha_{\ \mu\nu}$ and RST currents $j_{\alpha\mu}$ being specified by equations (\ref{2.19a}) and (\ref{2.26}), resp.

\indent In order to close the matter subdynamics, one has to specify now the gauge field dynamics. This is achieved by simply postulating the (non-Abelian) Maxwell-equations for the bundle connection $\MAmu$:
\begin{equation}\label{2.52}
\MDmo \MFmunu = -4\pi i \, \alpha_s \, \mathcal J_\nu \ .
\end{equation}
In component form, this equation reads
\begin{equation}\label{2.53}
\Dmo \Faomunu = 4\pi \, \alpha_s \, \jalonu \ ,
\end{equation}
provided one decomposes the (Hermitian) current operator $\mathcal J_\nu$, emerging on the right of the Maxwell equations (\ref{2.52}), in a similar way as was done for the curvature $\MFmunu$ (\ref{2.19a})-(\ref{2.19b}):
\begin{equation}\label{2.54}
\mathcal J_\mu = \exjmu \cdot \mathds 1 + i \, \jalomu \tau_\alpha \ .
\end{equation}
Now it is just this Maxwellian postulate (\ref{2.52}) which forces one to consider its consistency with the already existing RST dynamics for the spin-$\frac{1}{2}$ matter. The point here is that the generally valid bundle identity (\ref{2.23}), if being combined with the Maxwell equations (\ref{2.52}), immediately implies the following (non-Abelian) continuity equation for the current operator $\mathcal J_\mu$
\begin{equation}\label{2.55}
\D^\mu \mathcal J_\mu \equiv 0 \ ,
\end{equation}
which reads in component form
\begin{equation}\label{2.56}
D^\mu \, \jalomu \equiv 0 \ .
\end{equation}
\indent This looks very similar to the analogous continuity equation for the RST currents $\jalumu$ (\ref{2.27}), but it is by no means obvious that the latter current densities $\jalumu$ should be identical to the {\it "Maxwell currents"} $\jalomu$ building up the current operator $\mathcal J_\mu$ (\ref{2.54}). Thus, there arises the necessity of specifying the relationship between both currents $\jalomu$ (\ref{2.54}) and $\jalumu$ (\ref{2.26}). The solution of this {\it "compatibility problem"} consists in introducing the {\it compatibility tensor} $\Kaubu$ in such a way that it admits to pass over from one kind of current to the other so that the RST matter dynamics becomes consistent with the gauge field dynamics. More concretely we put
\begin{subequations}\label{2.57}
\begin{align}
\jalomu &= \Kaobo \, \jbumu \label{2.57a} \\
\jalumu &= \Kaubu \, \jbomo \label{2.57b} \\
(\Kaubu &\, \Kbogo = \delta^\gamma_{\ \alpha} ) \ . \nonumber
\end{align}
\end{subequations}

\indent Such a compatibility tensor $\Kaubu$ must own very special properties in order that it can unite the matter subdynamics and the gauge field subdynamics in a consistent way into the whole RST dynamics. First, observe that $\Kaubu$ must be covariantly constant
\begin{subequations}\label{2.58}
\begin{align}
&\Dmu \, \Kaubu = 0 \label{2.58a} \\
&\Dmu \, \Kaobo = 0 \ , \label{2.58b}
\end{align}
\end{subequations}
since both currents $\jalumu$ and $\jalomu$ have to obey simultaneously their corresponding continuity equations. Second, both currents transform under the adjoint representation of the N-particle structure group and therefore the compatibility tensor $\Kaubu$ must be invariant under such a transformation. Both requirements can be satisfied simultaneously by building up the compatibility object $\Kaubu$ by means of the generators $\tau_\alpha$ ($\alpha = 1 \dots N'^2$) of the reduced structure group as follows \cite{a43}:
\begin{equation}\label{2.59}
\Kaubu = c_1 \; \tr\left(\talu\right) \cdot \tr\left(\tbu\right) + c_2 \; \tr\left(\talu \cdot \tbu\right) \ ,
\end{equation}
where $c_1$ and $c_2$ are constants. Thus, the compatibility object $\mathcal K = \left\{ \Kaubu \right\}$ may be viewn also as a metric over the structure group, but not identical to the Killing-Cartan form.

\indent For arbitrary values of the constants $c_1$, $c_2$ the compatibility tensor $\mathcal K$ will admit non-vanishing diagonal elements $\{K_{\alpha\alpha}\}$ which implies the emergence of particle {\it self-interactions}. The point here is that for non-diagonal $\mathcal K$ the Maxwell currents $\jaomu$ ($a=1 \dots N$) are linear combinations of the RST currents $\jaumu$ which themselves are built up by the $a$-th particle field $\psi_a$ (as the $a$-th component of the N-particle wave function $\Psi$). Therefore anyone of the N fermions contributes to the $a$-th field strength $F^a_{\ \mu\nu}$ via the Maxwell equations (\ref{2.53}) which in turn acts back to the $a$-th particle via the internal Lorentz force (\ref{2.51}). Clearly, this mechanism establishes some kind of {\it self-force} of any particle upon itself; the corresponding {\it self-energy} will be discussed below. In order to present a non-trivial example, one may consider a system of two identical particles (e.g. the two electrons of the helium-like ions \cite{u5}), for which the structure group $\mathcal U(2)$ is four-dimensional and thus yields the following compatibility tensor $\mathcal K$:
\begin{equation}\label{2.60}
\Kaubu(u) = \eu \left(\begin{array}{cccc} \sinh u & -\cosh u & 0 & 0 \\ -\cosh u & \sinh u & 0 & 0 \\ 0 & 0 & 0 & -\eu \\ 0 & 0 & -\eu & 0 \end{array} \right) \ .
\end{equation}
This object depends no longer upon two constants $c_1, c_2$ (\ref{2.59}) but only on one {\it self-interaction parameter} $u$. The reason for this refers to the general requirement that the sum of the Maxwell currents $\jaomu$ ($a=1,2$) must equal the sum of the electromagnetic RST currents $\jaumu$ in order to yield the total current $j_\mu$:
\begin{equation}\label{2.61}
\sum_{a=1}^2 \, \jaomu = -\sum_{a=1}^2 \, \jaumu = -j_\mu \ .
\end{equation}
The particle self-interactions are described by the diagonal elements of $K_{ab} \; (\sim \sinh u)$ and therefore do vanish, if the self-interaction parameter tends to zero ($u \rightarrow 0$).

\section{Reduction of the Structure Group}

For the situation of N identical particles the mass operator $\mathcal M$ is proportional to the identitity
\begin{equation}\label{3.1}
\mathcal M \Rightarrow M \cdot \mathds 1
\end{equation}
and therefore the requirements (\ref{2.44})-(\ref{2.45}) do not represent restrictions for the structure algebra being spanned by the chosen $\mathfrak u(N)$ generators $\tau_\alpha$. (For the explicit form of the $3 \times 3=9$ three-particle generators $\{ \tau_\alpha \} =\{\tau_a, \chi_a, \bar \chi_a \}$ ($a=1 \dots 3$) see ref. \cite{a43}; and for the $2\times2 = 4$ two-particle generators $\{ \tau_\alpha \}= \{ \tau_a, \chi, \bar\chi \}$ ($a=1,2$) see ref. \cite{a35}.) Thus, for identical particles, which have the $(4N)$-dimensional $\Gamma_\mu$ (\ref{2.9}) as their total velocity operator, the full $\mathcal U (N)$ is active as the structure group. However when we consider, e.g., a three-particle system consisting of a positron and two electrons (or the negatively charged hydrogen ion) with the total velocity operator $\DGmu$ being given by the explicitly written terms in equation (\ref{2.8}), then we cannot expect that $\DGmu$ does still commute with all the nine $\mathcal U(3)$ generators $\tau_\alpha$ ($\alpha=1 \dots 9$).

\subsection{The Reduction Process}
Actually choosing these three-particle generators $\tau_\alpha$ as being specified in ref. \cite{a43}, one finds the following commutation relations
\begin{subequations}\label{3.2}
\begin{align}
&\left[\tau_a, \DGmu \right] = 0 \label{3.2a} \\
&\left[\chi_1, \DGmu \right] = 0 \label{3.2b} \\
&\left[\chi_2, \DGmu \right] = -2 \, \gamma_\mu \chi_2 \label{3.2c} \\
&\left[\chi_3, \DGmu \right] = 2 \, \gamma_\mu \chi_3 \label{3.2d} \ .
\end{align}
\end{subequations}
Therefore the original $\mathfrak u(3)$-valued connection $\MAmu$ (\ref{2.16})-(\ref{2.17c}) becomes restricted by the requirement (\ref{2.44}) to the five-dimensional subalgebra $\mathfrak u(2) \oplus \mathfrak u(1)$:
\begin{equation}\label{3.3}
\SMAmu \Rightarrow A^a_{\ \mu} \tau_a + \Bmu \chi_1 - \Bsmu \bar\chi_1 \doteqdot \emMAmu + \MBmu
\end{equation}
with the electromagnetic and exchange parts being evidently given by
\begin{subequations}\label{3.4}
\begin{align}
\emMAmu &= A^a_{\ \mu} \tau_a \equiv A^1_\mu \tau_1 + A^2_\mu \tau_2 + A^3_\mu \tau_3 \label{3.4a} \\
\MBmu &= \Bmu \chi_1 -\Bsmu \bar\chi_1 \ .
\end{align}
\end{subequations}
As a consequence, both covariant constancy conditions (\ref{2.42}) and (\ref{2.46}) are actually satisfied relative to the reduced connection $\SMAmu$ (\ref{3.3}).

\indent For the calculation of the curvature $\SMFmunu$ of the reduced connection $\SMAmu$, one needs the commutation relations for the five-dimensional structure group $\mathcal U(1) \times \mathcal U(2)$ which read \cite{a43}
\begin{subequations}\label{3.5}
\begin{align}
&\left[\tau_1, \chi_1 \right] = 0 \label{3.5a} \\
&\left[\tau_2, \chi_1 \right] = i \, \chi_1 \label{3.5b} \\
&\left[\tau_3, \chi_1 \right] = -i \, \chi_1 \label{3.5c} \\
&\left[\chi_1, \bar\chi_1\right] = i \, ( \tau_3 - \tau_2 ) \ , \label{3.5d}
\end{align}
\end{subequations}
and then the internal curvature is found to be of the form
\begin{equation}\label{3.6}
\SMFmunu = F^a_{\ \mu\nu} \tau_a + \Gmunu \chi_1 -\Gsmunu \bar \chi_1 \ ,
\end{equation}
with the curvature components being given in terms of the connection components as 
\begin{subequations}\label{3.7}
\begin{align}
\Femunu &= \nmu\Aenu - \nnu\Aemu \label{3.7a} \\
\Fzmunu &= \nmu\Aznu - \nnu\Azmu + i\left[\Bmu\Bsnu - \Bnu\Bsmu\right] \label{3.7b}\\
\Fdmunu &= \nmu\Adnu - \nnu\Admu - i\left[\Bmu\Bsnu - \Bnu\Bsmu\right] \label{3.7c}\\
\Gmunu  &= \nmu\Bnu - \nnu\Bmu + i\left[\Azmu - \Admu\right]\Bnu - i\left[\Aznu - \Adnu\right]\Bmu \label{3.7d} \ .
\end{align}
\end{subequations}

\indent Observe here that the first gauge field mode $\Femunu$ (\ref{3.7a}) is built up by the first vector potential $\Aemu$ in a completely Abelian way and therefore contains no exchange potential $B_\mu$! This important effect refers to our presumption that the first particle (i.e. positron or proton) is positively charged whereas the other two particles (i.e. the two electrons) are negatively charged. Thus, the very RST formalism automatically admits the exchange forces ($B_\mu$) to act exclusively among the {\it identical} particles (here the two electrons) but {\it not among different particles}. This nice feature of RST will become even more obvious for the matter field equations (see below) and has no counterpart in the conventional quantum theory, where the symmetrization or antisymmetrization of the wave functions of identical particles must be put in by hand (i.e. via the Pauli exclusion postulate \cite{b22}). The conventional theory would mathematically work also without applying the Pauli (anti-)symmetrization postulate, whereas the RST logic (\ref{2.44})-(\ref{2.46}) of conservation laws directly forbids the existence of exchange forces between non-identical particles. Thus, at least in this philosophical respect, RST is found to stand on a more fundamental level than the conventional theory.

\subsection{Reduced Maxwell Equations}

The reduction of the nine-dimensional structure group $\mathcal U(3)$ to the five-dimensional $\mathcal U(1) \times \mathcal U(2)$ must naturally leave its imprint on the gauge field dynamics. More concretely, it is not only the number of internal curvature components (\ref{3.6}), which is reduced from nine to five, but this reduction does apply also to the current operator $\mathcal J_\mu$ whose reduced internal part $\SMJmu$ (\ref{2.54}) reads now
\begin{equation}\label{3.8}
\SMJmu = i \left( \jaomu \tau_a + g_\mu \chi_1 - g^*_\mu \bar \chi_1 \right) \ ,
\end{equation}
where the first three current densities $\jaomu$ ($a=1,2,3$) are the real valued electromagnetic currents and $g_\mu$ ($\equiv j^4_{\ \mu}$) is the Maxwellian {\it exchange current}. Consequently, the reduced Maxwell equations (\ref{2.52}), or (\ref{2.53}), resp., appear now in explicit component form as follows:
\begin{subequations}\label{3.9}
\begin{align}
&\nmo\Femunu = \vpas \, \jeonu \label{3.9a} \\
&\nmo\Fzmunu + i\left[\Bmo\Gsmunu - \Bsmo\Gmunu\right] = \vpas \, \jzonu \label{3.9b} \\
&\nmo\Fdmunu - i\left[\Bmo\Gsmunu - \Bsmo\Gmunu\right] = \vpas \, \jdonu \label{3.9c} \\
&\nmo\Gmunu + i\left[\Azmu - \Admu\right]\Gmonu - i \left[\Fzmunu - \Fdmunu\right] \Bmo = \vpas \, g_\nu \ . \label{3.9d}
\end{align}
\end{subequations}

\indent Observe here again that the first gauge field mode $\Femunu$ obeys a strictly Abelian Maxwell equation because it is generated by the first (i.e. the non-identical) particle which cannot feel the exchange forces, in contrast to the two electrons (i.e. the second and third particles) which are identical and therefore generate cooperatively also an exchange potential $B_\mu$, cf. equations (\ref{3.9b})-(\ref{3.9d}). However it is also important to remark here that the total electromagnetic field strength $\Fmunu$ as the sum of the three elctromagnetic field modes
\begin{equation}\label{3.10}
\Fmunu = \sum\limits^3_{a=1}  \Famunu = \nabla_\mu A_\nu - \nabla_\nu A_\mu
\end{equation}
obeys an Abelian Maxwell equation
\begin{equation}\label{3.11}
\nmo \Fmunu = - \vpas \, \jnu
\end{equation}
with the total current $\jmu$ appearing as the sum of the three electromagnetic currents $\jaomu$:
\begin{equation}\label{3.12}
\sum\limits^3_{a=1} \jaomu = -\frac{1}{2}\sum\limits^3_{a=1} \jaumu = - \jmu \ .
\end{equation}
Subsequently we will see that these total objects $\left\{ \Fmunu, \jmu \right\}$, which characterize the three-particle system as a whole, do play an important role for the mechanism of self-interactions. Indeed, the latter kind of interactions will be revealed as a coupling of any individual particle $(a)$ to the total potential $A_\mu$ ($\doteqdot \sum\limits_a A^a_{\ \mu}$), to which itself contributes via its individual potential $A^a_{\ \mu}$.

\indent The Maxwell equations (\ref{3.9a})-(\ref{3.9d}) need a further completion as far as the Maxwell currents $\jalomu = \left\{ \jaomu ; g_\mu, g^*_\mu \right\}$ are concerned. Indeed it is necessary to elaborate their link to the wave function $\Psi$ in order that the RST matter and gauge field dynamics be closed. For this purpose, one has to clarify first the relationship between the RST currents $\jaumu$ (\ref{2.26}) and the components $\psi_a$ ($a=1,2,3$) of the three-particle wave function $\Psi = \left\{ \psi_1, \psi_2, \psi_3 \right\}$. By some simple mathematics, one arrives at the following relations:
\begin{subequations} \label{3.13}
\begin{align}
\jeumu \doteqdot \bar \Psi \upsilon_{1\mu} \Psi &= \ \bar \psi_2 \gamma_\mu \psi_2 + \bar \psi_3 \gamma_\mu \psi_3 \ \doteqdot \ k_{2\mu} + k_{3\mu} \label{3.13a} \\
\jzumu \doteqdot \bar \Psi \upsilon_{2\mu} \Psi &= - \bar \psi_1 \gamma_\mu \psi_1 + \bar \psi_3 \gamma_\mu \psi_3 \doteqdot -k_{1\mu} + k_{3\mu} \label{3.13b} \\
\jdumu \doteqdot \bar \Psi \upsilon_{3\mu} \Psi &= - \bar \psi_1 \gamma_\mu \psi_1 + \bar \psi_2 \gamma_\mu \psi_2 \doteqdot -k_{1\mu} + k_{2\mu}  \ . \label{3.13c}
\end{align}
\end{subequations}
Here we have introduced the {\it Dirac currents} $k_{a\mu}$ in a self-evident way; and analogously one introduces the RST exchange current $h_\mu$ through
\begin{subequations} \label{3.14}
\begin{align}
\jvumu &\equiv h_\mu \doteqdot \bar \psi_2 \gamma_\mu \psi_3 \label{3.14a} \\
\jfumu &\equiv - h^*_\mu \ = -\bar\psi_3 \gamma_\mu \psi_2 . \label{3.14b}
\end{align}
\end{subequations}
It is useful for the subsequent discussions to reexpress the Dirac currents $\kaumu$ also in terms of the RST currents $\jaumu$:
\begin{subequations} \label{3.15}
\begin{align}
\keumu &= \frac{1}{2} \left( \jeumu - \jzumu - \jdumu \right) \label{3.15a} \\
\kzumu &= \frac{1}{2} \left( \jeumu - \jzumu + \jdumu \right) \label{3.15b} \\
\kdumu &= \frac{1}{2} \left( \jeumu + \jzumu - \jdumu \right) \ . \label{3.15c}
\end{align}
\end{subequations}
The total current $\jmu$ (\ref{2.10})-(\ref{2.11}) is recovered then simply as the "charge-weighted" sum of those Dirac currents, i.e.
\begin{equation}\label{3.16}
\jmu = - \keumu + \kzumu + \kdumu \ .
\end{equation}
This pleasent result arises from the fact that the gauge velocity operators $\upsilon_{a\mu}$ (\ref{2.24}) are chosen in such a way that their sum just yields the total velocity operator $\DGmu$ (\ref{2.8}):
\begin{equation}\label{3.17}
\sum^3_{a=1} \upsilon_{a\mu} = i \Big( \sum^3_{a=1} \tau_a \Big)  \DGmu = 2 \, \DGmu \ .
\end{equation}
Consequently the sum of the RST currents $\jaumu$ just yields the total current $\jmu$ (up to a factor of 2)
\begin{equation}\label{3.18}
\sum^3_{a=1} \jaumu = \sum^3_{a=1} \bar \Psi \upsilon_{a\mu} \Psi = \bar \Psi \Big( \sum^3_{a=1} \upsilon_{a\mu} \Big)\!\Psi = 2 \, \bar \Psi \DGmu \Psi = 2 \, \jmu \ .
\end{equation}
which is implied also by equations (\ref{3.13a})-(\ref{3.13c}) and moreover verifies the former assertion (\ref{3.12}).

\indent Since the RST currents $\jaumu$ can be represented also in terms of the Dirac currents $\left\{\kaumu, \hmu \right\}$, it must be possible to reformulate the (non-Abelian) source equations (\ref{2.27}) for the RST currents also in terms of those Dirac currents. Clearly, this system of source equations must appear again as the reduced form of the original source system for three identical particles \cite{a43}. Therefore it is very instructive to reexamine here once more how this reduction of the source equations comes about. To this end, one carries out the covariant differentiation process for the RST currents $\jalumu$ under extensive use of the RST kinematics and dynamics being described above, which then finally yields
\begin{equation}\label{3.19}
D^\mu \jalumu = - \frac{1}{\hbar c} \tr \left\{ \mathcal I \cdot \left[\mathcal M c^2, \tau_\alpha \right] \right\} \ .
\end{equation}
Thus for {\it identical particles} (\ref{3.1}), it is no problem to deduce from here the source equations (\ref{2.27}) because the commutator of the mass operator $\mathcal M$ and the original structure algebra $\{\tau_\alpha\}$ vanishes trivially. However, if one of the three particle masses is different from the other two, e.g.
\begin{equation}\label{3.20}
\mathcal M = \left(\begin{array}{ccc} M_p \cdot \mathbf 1 & 0 & 0 \\ 0 & M_e \cdot \mathbf 1 & 0 \\ 0 & 0 & M_e \cdot \mathbf 1 \end{array} \right) \ ,
\end{equation}
then the requirement of vanishing commutator (\ref{2.44}) yields a non-trivial reduction of the structure algebra, as was described above. The corresponding reduced source equations read in terms of the Dirac currents:
\begin{subequations} \label{3.21}
\begin{align}
\nmo \keumu &= 0 \label{3.21a} \\
\nmo \kzumu &= i \left[ \Bmo \hmu- \Bsmo \hsmu \right] \label{3.21b} \\
\nmo \kdumu &= - i \left[ \Bmo \hmu- \Bsmo \hsmu \right] \label{3.21c} \\
\nmo \hmu &= i \left[ \Azmu - \Admu \right]\hmo + i \, \Bsmo \left[\kzumu -\kdumu \right] \ . \label{3.21d}
\end{align}
\end{subequations}

\indent These source equations represent the component form of the operator equation (\ref{2.55}) and therefore are the integrability conditions for the non-Abelian Maxwell equations (\ref{2.52}). On the other hand, these source equations (\ref{3.21a}) -(\ref{3.21d}) are the direct implication of the RST matter dynamics, and therefore we have thus demonstrated that in the case of pure states the RST matter dynamics ownes the status of a sufficient integrability condition for the non-Abelian gauge field-equations (\ref{2.52})! In this way, the compatibility of the matter and gauge field dynamics becomes evident once more. (See also the discussion of integrability conditions in ref.\cite{a43}).

\indent Finally, the Maxwell currents $\{\jalomu \} = \{\jaomu; g_\mu \}$ must be expressed in terms of the Dirac currents $\{k_{a\mu}; h_\mu\}$ in order that the coupling of the Maxwell equations (\ref{3.9a})-(\ref{3.9d}) to the matter fields $\psi_a$ becomes manifest and the hole RST system becomes closed. To this end, one simply rewrites the RST currents $\jalumu$ in terms of the Dirac currents, cf. (\ref{3.14a})-(\ref{3.15c}), and substitutes this into the link (\ref{2.57a}) between the Maxwell and RST currents. However, this procedure requires to first specify the five-dimensional fibre metric $K_{\alpha\beta}$ from its general shape (\ref{2.59}). Applying here a similar reasoning as for the situation with two identical particles (\ref{2.60})-(\ref{2.61}), fixes the constants $c_1, c_2$ in equation (\ref{2.59}) to
\begin{subequations}\label{3.22}
\begin{align}
c_1 &= \frac{1}{32} \eu \left( \cosh{u} + \sinh{u} \right) \label{3.22a}\\
c_2 &= - \frac{1}{4} \eu \left( \cosh{u} + 2 \sinh{u} \right) \ , \label{3.22b}
\end{align}
\end{subequations}
and correspondingly the fibre metric $\Kaubu$ must look as follows
\begin{equation}\label{3.23}
\Kaubu(u) = \footnotesize{ \eu \left(\begin{array}{ccccc} 2\sinh u & -\cosh u & -\cosh u & 0 & 0 \\ -\cosh u & 2\sinh u & -\cosh u & 0 & 0 \\ -\cosh u & -\cosh u & 2\sinh u & 0 & 0 \\ 0 & 0 & 0 & 0 & -\cosh u -2 \sinh u  \\ 0 & 0 & 0 & -\cosh u -2 \sinh u & 0 \end{array}  \right)} \ .
\end{equation}
On priciple, this is the same form as for the two-particle situation (\ref{2.60}), apart from the substitution $\sinh u \Rightarrow 2 \sinh u$. This substitution implies that the strength of self-coupling of three-particles (i.e. the diagonal elements of $\Kaubu$) is twice as much as for the two-particle situation, provided the strength of the mutual interaction (measured by the off-diagoanl elements, $\cosh u$) remains the same.

\indent But now that the fibre metric $\Kaubu$ for the reduced structure group is fixed, it is possible to explicitly perform the transition (\ref{2.57a}) to the Maxwell currents $\jalomu$ which are needed to solve the Maxwell equations (\ref{2.52}). For the sake of convenience, one introduces the following self-interaction parameters:
\begin{subequations}\label{3.24}
\begin{align}
T_p(u) \doteqdot \frac{1-\tanh u}{1+2\tanh u} \label{3.24a} \\
T_s(u) \doteqdot \frac{\tanh u}{1+2\tanh u} \ , \label{3.24b}
\end{align}
\end{subequations}
and then the Maxwell currents $\jalomu$ read in terms of their Dirac counterparts
\begin{subequations}\label{3.25}
\begin{align}
\jeomu &= T_p(u) \cdot \keumu - T_s(u) \cdot \jmu \label{3.25a} \\
\jzomu &= - T_p(u) \cdot \kzumu - T_s(u) \cdot \jmu \label{3.25b} \\
\jdomu &= - T_p(u) \cdot \kdumu - T_s(u) \cdot \jmu \label{3.25c} \\
\jvomu &\equiv g_\mu = T_p(u) \cdot \hsmu \label{3.25d} \\
\jfomu &\equiv - g^*_\mu = -T_p(u) \cdot \hmu \ . \label{3.25e}
\end{align}
\end{subequations}
In order to gain some confidence in this transformation, one easily checks the three-particle analogue (\ref{3.12}) of the two-particle case (\ref{2.61}) which is most easily seen to be verified with the help of
\begin{equation}\label{3.26}
T_p(u) + 3 \, T_s(u) = 1 \ ,
\end{equation}
being an immediate consequence of the notations (\ref{3.24a})-(\ref{3.24b}).

\indent However the most interesting point with the Maxwell currents $\jaomu$ ($a=1 \dots 3$) refers to the fact that they are built up also by the total current $\jmu$ besides the expected emergence of the Dirac currents $\kaumu$! From this reason, anyone of the electromagnetic gauge field modes $\Famunu$ contains a constituent which is excited by the total current $\jmu$, besides the main contribution coming from the $a$-th Dirac current $\kaumu$ according to the Maxwell equations (\ref{3.9a})-(\ref{3.9d}). Clearly this is nothing else than a very specific kind of particle self-interaction because the internal Lorentz force $\Sfnu(\mathbf a)$ (\ref{2.51}) upon the $\mathbf a$-th particle
\begin{equation}\label{3.27}
\Sfnu(\mathbf a) = \hbar c \sum^3_{b=1} K_{\mathbf a b} F^{\mathbf a}_{\ \mu\nu} j^{b \mu} 
\end{equation}
contains a term which originates from its own field $F^{\mathbf a}_{\ \mu\nu}$ ! A further consequence of these self-interactions is the self-energy shift of the energy levels of the bound systems (to be readily considered in detail). Observe however that the strength of these self-interactions is determined by the value of the self-interaction parameter $u$, so that for $u \rightarrow 0$ the Maxwell currents $\jalomu$ (\ref{3.25a})-(\ref{3.25e}) coincide with the Dirac currents and the self-interaction vanishes.

\subsection{Reduced Matter Dynamics}
\indent Naturally, the reduction of the structure group $\mathcal U(3) \Rightarrow \mathcal U(1) \times \mathcal U(2)$ must also imply certain effects upon the matter dynamics. This may be seen most clearly by reconsidering the Dirac equation (\ref{2.49}) in component form. In the compact form, the gauge covariant derivative of the three-particle wave function $\Psi$ is as usual
\begin{equation}\label{3.28}
\MDmu\Psi = \pmu\Psi + \MAmu\Psi \ .
\end{equation}
But if the wave function is resolved here into components
\begin{equation}\label{3.29}
\Psi = \left(\begin{array}{c} \psi_1 \\ \psi_2 \\ \psi_3 \end{array} \right) \ ,
\end{equation}
the covariant derivative (\ref{3.28}) appears as
\begin{equation}\label{3.30}
\MDmu\Psi = \left(\begin{array}{c} D_\mu \psi_1 \\ D_\mu \psi_2 \\ D_\mu \psi_3 \end{array} \right)
\end{equation}
with the derivatives of the components $\psi_a$ being formed by means of the reduced connection $\SMAmu$ (\ref{3.3}) in the following way:
\begin{subequations}\label{3.31}
\begin{align}
\Dmu \psi_1 &= \pmu\psi_1 - i \left[\exAmu + \Azmu + \Admu \right] \psi_1 \label{3.31a} \\
\Dmu \psi_2 &= \pmu\psi_2 - i \left[\exAmu + \Aemu + \Admu \right] \psi_2 - i \Bmu \psi_3 \label{3.31b} \\
\Dmu \psi_3 &= \pmu\psi_3 - i \left[\exAmu + \Aemu + \Azmu \right] \psi_3 - i \Bsmu \psi_2 \ . \label{3.31c}
\end{align}
\end{subequations}

As expected, anyone of the three particles feels the electromagnetic potential $\Amu$ being generated by the other two particles via the Maxwell equations (\ref{3.9a})-(\ref{3.9c}). However, the second and third particles (i.e. the two identical electrons) do additionally feel also the exchange potential $\Bmu$ which is generated by the cooperatively produced exchange current $h_\mu$ (\ref{3.14a}) according to the last Maxwell equation (\ref{3.9d}). In contrast to this, the first particle (\ref{3.31a}), i.e. the positron or proton, undergoes exclusively the electromagnetic interactions mediated by the potentials $A^a_{\ \mu}$; $a=2,3$. This discrimination of non-identical particles is in excellent agreement with the experimental evidence and appears here as an immediate consequence of the reduction process which itself originates from the intrinsic logic of RST. Thus, the status of the exchange interactions is somewhat different in RST and in the conventional theory: Whereas in the latter framework the existence or non-existence of exchange effects arises as an implication of the kinematical postulate of wave function (anti)symmetrization (or (anti)commutation of field operators, resp.), the presence or absence of exchange effects originates in RST through the implications of the conservation laws!

\indent For the concrete calculations, the abstract three-particle Dirac equation (\ref{2.49}) with the mass operator $\mathcal M$ (\ref{3.20}) will be resolved into its components
\begin{subequations}\label{3.32}
\begin{align}
&i \hbar \, \gamma^\mu D_\mu \psi_1 = - M_p c \psi_1 \label{3.32a} \\
&i \hbar \, \gamma^\mu D_\mu \psi_2 = M_e c \psi_2 \label{3.32b} \\
&i \hbar \, \gamma^\mu D_\mu \psi_3 = M_e c \psi_3 \label{3.32c} \ ,
\end{align}
\end{subequations}
where the covariant derivatives are specified through equations (\ref{3.31a})-(\ref{3.31c}). Observe that the mass $M_p$ of the positively charged(i.e. first) particle enters the Dirac system (\ref{3.32a})-(\ref{3.32c}) with a minus sign, but this does not spoil the consistency of the energy eigenvalue problem for the three-particle system (see below for the discussion of the energy functional of the bound states).

\section{Stationary Bound States}
In order to test the practical usefulness of RST, one may consider the (experimentally available) energy levels of stationary bound systems, e.g. the negatively charged hydrogen or positronium ion \cite{a44,a45,a46}. Such three-particle systems provide us with the additional advantage of leaving apart one of the two electrons so that one ends up with the neutral positronium or hydrogenium. This cutting of a three-particle system down to a two-particle system will yield further insight into the RST mechanism of the self-interactions (separate paper). However subsequently, we restrict ourselves to elaborating the general form of the three-particle eigenvalue system for determining the energy levels of the bound systems.

\indent The calculation of the desired energy levels will consist in a three-step procedure:
\begin{enumerate}
\item one has to set up the coupled system of {\it mass eigenvalue equations} to be deduced from the Dirac system (\ref{3.32a})-(\ref{3.32c}),
\item since these mass eigenvalue equations contain the coupling of the wave functions $\psi_a$ to the gauge fields, one has to complement this eigenvalue system by supplying also the (non-Abelian) Poisson equations for the gauge potentials which must be deduced from the original Maxwell equations (\ref{2.52}),
\item once the solutions of this coupled Dirac-Poisson system have been obtained, one takes the value of the RST energy functional $E_T$ upon these solutions and thus ends up with the desired energy levels. The latter may then be used in order to calculate further observable quantities such as hyperfine splittings etc.
\end{enumerate}
Such a procedure has been applied successfully to systems of identical particles \cite{a43,u5}; but the point here is now that one must be able to demonstrate that this scheme does work also for systems which contain both identical and non-identical particles! Subsequently we will exemplify the general situation by resorting to the presently considered three-particle system. This will yield the result that RST can treat a mixture of identical and non-identical particles in a completely consistent way.

\subsection{Mass Eigenvalue Equations}
For the stationary bound states $\psi_a(\vec{r},t)$, to be considered from now on, one tries the usual product ansatz ($a=1,2,3$)
\begin{equation}\label{4.1}
\psi_a(\vec{r},t) = \exp \left[-i\frac{M_ac^2 \, t}{\hbar} \right] \cdot \psi_a(\vec{r}) \ .
\end{equation}
The corresponding mass eigenvalue equations for the determination of the {\it mass eigenvalues} $M_a$ arise simply by substituting the present ansatz (\ref{4.1}) into the coupled Dirac equations (\ref{3.32a})-(\ref{3.32c}). The goal is here to verify that the different signs in front of the rest masses $M_p$, $M_e$ of the positively and negatively charged particles do {\it not} induce any inconsistency into the energy eigenvalue problem.

\indent The mass eigenvalue equation for the first (i.e. positively charged) particle is deduced from the general Dirac equation (\ref{3.32a}) by means of the present ansatz (\ref{4.1}) as
\begin{multline}\label{4.2}
i \, \vec \gamma \bdot \vec \nabla \, \psi_1(\vec r) + [\exAo + \zAo + \dAo] \, \gamma^0 \, \psi_1(\vec r) - \\ - [\vec A_{ex} + \vec A_2 + \vec A_3 ] \bdot \vec \gamma \; \psi_1(\vec r) = - \frac{M_p + \gamma^0 \, M_1}{\hbar}c \cdot \psi_1(\vec{r}) \ .
\end{multline}
Here both the external potentials $\exAmu$ and the internal potential $A^a_{\ \mu}$ are assumed to be time-independent on account of the stationary situation, i.e. we put
\begin{subequations}\label{4.3}
\begin{align}
\exAmu &\Rightarrow \{ \exAo(\vec{r}); - \vec A_{ex}(\vec r) \} \label{4.3a}\\
A^a_{\ \mu} &\Rightarrow \{ \aAo(\vec{r}); - \vec A_a(\vec r) \} \label{4.3b}\ .
\end{align}
\end{subequations}
Thus, the eigenvalue equation (\ref{4.2}) for the positively charged particle does not contain the exchange potential $B_\mu$ which just meets with the physical requirement that the positive charge cannot undergo exchange interactions with the negative charges. However, such exchange forces are felt by the two electrons ($a=2,3$) whose mass eigenvalue equations appear from (\ref{3.32b})-(\ref{3.32c}) as
\begin{subequations}\label{4.4}
\begin{multline}\label{4.4a}
i \, \vec \gamma \bdot \vec \nabla \, \psi_2(\vec r) + [\exAo + \eAo + \dAo] \, \gamma^0 \, \psi_2(\vec r) + B_0(\vec r) \, \gamma^0 \, \psi_3(\vec r) - \\ - [\vec A_{ex} + \vec A_1 + \vec A_3 ] \bdot \vec \gamma \; \psi_2(\vec r) - \vec B (\vec r) \bdot \vec \gamma \; \psi_3(\vec r) = \frac{M_e - \gamma^0 \, M_2 }{\hbar}c \cdot \psi_2(\vec{r})
\end{multline}
\begin{multline}\label{4.4b}
i \, \vec \gamma \bdot \vec \nabla \, \psi_3(\vec r) + [\exAo + \eAo + \zAo] \, \gamma^0 \, \psi_3(\vec r) + B^*_0(\vec r) \, \gamma^0 \, \psi_2(\vec r) - \\ - [\vec A_{ex} + \vec A_1 + \vec A_2 ] \bdot \vec \gamma \; \psi_3(\vec r) - \vec B^* (\vec r) \bdot \vec \gamma \; \psi_2(\vec r) = \frac{M_e - \gamma^0 \, M_3}{\hbar}c \cdot \psi_3(\vec{r}) \ .
\end{multline}
\end{subequations}
Indeed, both electrons are seen to be subjected not only to the usual electromagnetic forces, mediated by the electromagnetic potentials $\exAmu$ and $A^a_{\ \mu}$, but obviously they do feel also the exchange potential $B_\mu$! It is true, the latter potential is not time-independent as its electromagnetic counterparts (\ref{4.3a})-(\ref{4.3b}); however the time-dependence is found to be of the following form:
\begin{equation}\label{4.5}
B_\mu (\vec r, t) = \exp \left[-i \frac{M_2-M_3}{\hbar} c^2 \, t \right] \cdot B_\mu(\vec r) \ ,
\end{equation}
so that the time factor cancels for the two electronic mass eigenvalue equations (\ref{4.4a})-(\ref{4.4b}). Thus the question of exchange forces for (non-)identical particles appears to be clarified in a satisfactory way; but simultaneously there do arise other nearby questions of consistency, namely:
\begin{enumerate}
\item the rest mass $M_p$ of the positively charged particle enters its eigenvalue equation (\ref{4.2}) with a different sign in comparison to the electron's rest mass $M_e$ in equations (\ref{4.4a})-(\ref{4.4b}) and
\item the electromagnetic potentials $\exAmu$, $A^a_{\ \mu}$  do enter the eigenvalue equations of all three particles with the {\it same} sign, despite the fact, that the positive and negative charges do react {\it oppositely} to the same electric field!
\end{enumerate}
However these apparent contradictions can be clarified easily by considering the {\it mass functionals} $M_a[\Psi]$. Indeed these functionals are built up by the energies of kinetic, potential and rest-mass type, resp., so that one can explicitly check the signs of the various contributions.

\subsection{Mass Functional}
\indent The general structure of the preceding mass eigenvalue equations (\ref{4.2}) and (\ref{4.4a})-(\ref{4.4b}) admits to express the mass eigenvalues $M_a$ in terms of the wave functions $\psi_a$ and potentials $A^a_{\ \mu}$, $B_\mu$. This will yield more insight into the individual contributions to the total energy $E_T$. The desired mass functionals are easily obtained by contracting the corresponding mass eigenvalue equations with $\bar \psi_a(\vec r)$ and integrating over whole three-space ($t=\textnormal{const.}$). In this way, the mass functional for the positively charged particle is deduced from its eigenvalue equation (\ref{4.2}) as
\begin{equation}\label{4.6}
-\hat z_1 \cdot M_1 c^2 = \mathcal Z^2_{(\mathbf 1)} \cdot M_p c^2 + 2 \, T_{kin(\mathbf 1)} + \hat z_1 \cdot (M^{(es)}_{1,e}c^2 + M_I^{(e)} c^2 + M_I^{(m)}c^2 ) \ .
\end{equation}
The striking point with this result is that the first mass eigenvalue $M_1$ is negative since the first term on the right-hand side is positive and dominates all the other terms which will now be explained in detail.

\indent First, turn to the pre-factor $\hat z_1$ of the mass eigenvalue $M_1c^2$ which is defined as follows ($a=1,2,3$):
\begin{equation}\label{4.7}
\hat z_a \doteqdot \int d^3\vec r \; \ako(\vec r) \ .
\end{equation}
Here the Dirac currents $k_{a\mu}$ (\ref{3.13a})-(\ref{3.13c}) are assumed to be time-independent, just as the corresponding potentials $A^a_{\ \mu}$ (\ref{4.3b})
\begin{equation}\label{4.8}
k_{a\mu} (\vec r, t) = \left\{\ako(\vec r); - \vec k_a(\vec r) \right\} \ ,
\end{equation}
with the current components reading in terms of the wave functions $\psi_a(\vec r)$
\begin{subequations}\label{4.9}
\begin{align}
\ako(\vec r) &\doteqdot \bar \psi_a(\vec r) \gamma_0 \, \psi_a(\vec r) \equiv \psi_a^\dagger(\vec r) \cdot \psi_a(\vec r) \label{4.9a} \\
\vec k_a(\vec r) &\doteqdot \bar \psi_a(\vec r) \vec \gamma \, \psi_a(\vec r) \label{4.9b} \ .
\end{align}
\end{subequations}
Thus, the relation (\ref{4.7}), in combination with the density ${}^{(a)}\!k_0$ (\ref{4.9a}), is very similar to the usual normalization condition for the wave functions $\psi_a$; but in general the normalization parameter $\hat z_a$ is slightly different from unity ($\hat z_a \neq 1$) because the densities $\ako(\vec r)$ must be slightly modified in the presence of non-trivial exchange fields in order to yield a modified density $\alo(\vec r)$ which then undergoes the exact normalization condition ($a=1,2,3$)
\begin{equation}\label{4.10}
1 = \int d^3\vec r \; \alo(\vec r) \ ,
\end{equation}
see ref. \cite{a43}. Nevertheless, since the exchange effect is relatively small, the value of the normalization parameters $\hat z_a$ will be found in most situations to be close to unity.

\indent Next, consider the first term on the right of equation (\ref{4.6}) which essentially represents the rest-mass energy $M_pc^2$ of the first particle, where this kind of energy dominates all the other energy contributions (i.e. kinetic and potential energies of electric ($e$) and magnetic ($m$) origin). However it is important to observe that the rest mass energy ($M_pc^2$) becomes slightly modified by the {\it renormalization constant} $\mathcal Z^2_{(\bf 1)}$
\begin{equation}\label{4.11}
\mathcal Z^2_{(\bf a)} \doteqdot \int d^3\vec r \; \bar \psi_a(\vec r) \psi_a(\vec r) \ .
\end{equation}
It has been shown in a previous paper \cite{a35} that the presence of such a renormalization constant $\mathcal Z^2_{(\bf a)}$ is necessary in order to compensate for the {\it doubling} of the kinetic energy $T_{kin(\bf 1)}$ in (\ref{4.6}):
\begin{equation}\label{4.12}
T_{kin(\bf 1)} \doteqdot \frac{i}{2} \, \hbar c \int d^3\vecr \; \bar \psi_1 \vec \gamma \bdot \vec \nabla \psi_1 \ ,
\end{equation}
see below for the non-relativistic limit of $T_{kin(\bf a)}$.

\indent Finally, the mass equivalents of the various potential energies have to be specified. The electrostatic interaction energy of the first particle with the external source ($\leadsto \exAo(\vec r)$) is defined by
\begin{equation}\label{4.13}
\hat z_1 \cdot M^{(es)}_{1,e}c^2 \doteqdot \hbar c \int d^3\vec r \; \exAo (\vec r) \cdot \eko(\vec r) \ .
\end{equation}
Similarly, the electrostatic interaction energy of the first particle with the two electrons is given by
\begin{equation}\label{4.14}
\hat z_1 \cdot M_I^{(e)}c^2 \doteqdot \hbar c \int d^3\vec r \; [\zAo(\vec r) + \dAo(\vec r) ] \cdot \eko(\vec r) \ ,
\end{equation}
and analogously for the magnetostatic (i.e. spin-spin) interaction energy
\begin{equation}\label{4.15}
\hat z_1 \cdot M_I^{(m)} c^2 \doteqdot -\hbar c \int d^3\vec r \; [\vec A_2(\vec r) + \vec A_3(\vec r) ] \bdot \vec k_1(\vec r)
\end{equation}
(for the sake of simplicity, the external source is assumed to emit no magnetic field, i.e. we put $\vec A_{ex}(\vec r) \equiv 0$).

\indent Now, concerning the raised consistency question, it may appear somewhat suspicious that the first mass eigenvalue $M_1c^2$ (\ref{4.6}) should be negative. However, as we will readily see, the mass eigenvalues $M_ac^2$ are not identical to the total energy $E_T$ of the RST field configuration; and actually we will find that the {\it negative} mass eigenvalue $M_1c^2$ contributes a {\it positive} amount to the total field energy $E_T$. But apart from this question, there is a further point of concern; namely the interaction energy of a {\it positively} charged particle with a given electric potential (e.g. $\exAo(\vec r)$) is minus the energy of a {\it negatively} charged particle in the same external field $\exAo(\vec r)$. Therefore, deducing now the electronic mass functionals $M_ac^2[\Psi]$ ($a=2,3$) from the corresponding eigenvalue equations (\ref{4.4a})-(\ref{4.4b}) provides us with a test about whether the positivity and negativity of the electric charges is correctly incorporated in RST.

\indent The desired electronic mass functionals are found to be of the following form:
\begin{subequations}\label{4.16}
\begin{multline}\label{4.16a}
\hat z_2 \cdot M_2c^2 = \mathcal Z^2_{(\bf 2)} \cdot M_ec^2 + 2\, T_{kin(\bf 2)} + \\ + \hat z_2 \cdot (M_{2,e}^{(es)}c^2 + M_{I\!I}^{(e)} c^2 + M_{I\!I}^{(m)} c^2 + M_h c^2 + M_g c^2)
\end{multline}
\begin{multline}\label{4.16b}
\hat z_3 \cdot M_3 c^2 = \mathcal Z^2_{(\bf 3)} \cdot M_ec^2 + 2\, T_{kin(\bf 3)} + \\ + \hat z_3 \cdot (M_{3,e}^{(es)}c^2 + M_{I\!I\!I}^{(e)} c^2 + M_{I\!I\!I}^{(m)} c^2 + M_h^* c^2 + M_g^* c^2 ) \ .
\end{multline}
\end{subequations}
Here, the mass equivalents $M_{a,e}^{(es)}c^2$ of the external interaction energy are given by ($a=2,3$)
\begin{equation}\label{4.17}
\hat z_a \cdot M^{(es)}_{a,e} c^2= -\hbar c \int d^3\vec r \; \exAo(\vec r) \cdot \ako(\vec r)
\end{equation}
and thus have just the opposite sign as for the positively charged particle (\ref{4.13}). Thus, at least with respect to the external interactions, the charge dichotomy is correctly accounted for.

\indent Next, one finds for the interparticle interaction energy of the electric type:
\begin{subequations}\label{4.18}
\begin{align}
\hat z_2 \cdot M_{I\!I}^{(e)}c^2 &= -\hbar c \int d^3\vec r \; [\eAo(\vec r) + \dAo(\vec r) ] \cdot \zko(\vec r) \label{4.18a} \\
\hat z_3 \cdot M_{I\!I\!I}^{(e)}c^2 &= -\hbar c \int d^3\vec r \; [\eAo(\vec r) + \zAo(\vec r) ] \cdot \dko(\vec r) \label{4.18b} \ .
\end{align}
\end{subequations}
Now look here at anyone of the three electrostatic potentials $\aAo(\vec r)$ ($a=1,2,3$) and check the signs of the interaction energies with respect to the other two particles $b$ ($b\neq a$):
\begin{enumerate}
\item the first potential $\eAo(\vec r)$ (emitted by the positive charge) implies the same signs for the interactions with the two electrons, cf. (\ref{4.18a})-(\ref{4.18b});
\item the second potential $\zAo(\vec r)$ implies the emergence of different signs with respect to the interaction with an electron (\ref{4.18b}) or with respect to a positron (\ref{4.14}), etc.
\end{enumerate}
\indent The same logic does apply also with respect to the magnetic interactions of both electrons:
\begin{subequations}\label{4.19}
\begin{align}
\hat z_2 \cdot M_{I\!I}^{(m)}c^2 &= \hbar c \int d^3\vec r \; [\vec A_1(\vec r) + \vec A_3(\vec r) ] \bdot \vec k_2(\vec r) \label{4.19a} \\
\hat z_3 \cdot M_{I\!I\!I}^{(m)}c^2 &= \hbar c \int d^3\vec r \; [\vec A_1(\vec r) + \vec A_2(\vec r) ] \bdot \vec k_3(\vec r) \label{4.19b} \ .
\end{align}
\end{subequations}
For this magnetic case it is necessary to observe that (for vanishing self-interactions, $u=0$, for the sake of simplicity) the Maxwell currents $\vec j_a$ (\ref{3.25a})-(\ref{3.25c}) are arranged in a different way relative to the Dirac currents $\vec k_a$:
\begin{subequations}\label{4.20}
\begin{align}
\vec j_1 &= \vec k_1 \label{4.20a} \\
\vec j_2 &= -\vec k_2 \label{4.20b} \\
\vec j_3 &= -\vec k_3 \label{4.20c} \ .
\end{align}
\end{subequations}
If this circumstance is respected, one finds for the magnetic interparticle interactions the same logical arrangement as was described above for the electric interparticle interactions.

\indent Summarizing, one finds that the electromagnetic interparticle interactions of different charges are correctly incorporated in the RST formalism. Furthermore, also the exchange interactions appear at the right places: no exchange contribution to the mass functional $M_1c^2$ (\ref{4.6}) of the single positive charge, but non-trivial exchange contributions to the electronic mass functionals $M_2c^2$, $M_3c^2$ (\ref{4.16a})-(\ref{4.16b}):
\begin{subequations}\label{4.21}
\begin{align}
\hat z_2 \cdot M_h c^2 &= - \hbar c \int d^3\vec r \; B_0(\vec r) \cdot h_0(\vec r) \label{4.21a}\\
\hat z_2 \cdot M_g c^2 &=   \hbar c \int d^3\vec r \; \vec B(\vec r) \bdot \vec h(\vec r) \label{4.21b} \ \textnormal{, etc.}
\end{align}
\end{subequations}

\indent Despite this positive outcome, there remains to be clarified one peculiarity in connection with the mass functionals, and this refers to the kinetic energies $T_{kin(\bf a)}$ of the three particles, which has not yet been inspected:
\begin{subequations}\label{4.22}
\begin{align}
T_{kin(\bf 1)} &= \frac{i}{2} \, \hbar c \int d^3\vec r \; \bar \psi_1(\vec r) \vec \gamma \bdot \vec \nabla \, \psi_1(\vec r) \label{4.22a} \\
T_{kin(\bf 2)} &= -\frac{i}{2} \, \hbar c \int d^3\vec r \; \bar \psi_2(\vec r) \vec \gamma \bdot \vec \nabla \, \psi_2(\vec r) \label{4.22b} \\
T_{kin(\bf 3)} &= -\frac{i}{2} \, \hbar c \int d^3\vec r \; \bar \psi_3(\vec r) \vec \gamma \bdot \vec \nabla \, \psi_3(\vec r) \label{4.22c}  \ .
\end{align}
\end{subequations}
Evidently, one becomes faced here with the question why positive and negative charges must have different signs for the definition of their kinetic energies? In order to clarify this question , it is very instructive to consider the non-relativistic limit of the kinetic energy.

\subsection{Non-relativistic Approximation}
As is well-known from the one-particle theory, the four components of the Dirac wave functions $\psi_a$ are not of equal magnitude but may differ by a factor of $\alpha_{\rm S}^2$ \cite{b11}. Therefore it is adequate to split up the four-component Dirac spinors $\psi_a$ into the "upper" ($\varphi_+$) and "lower" ($\varphi_-$) Pauli components
\begin{equation}\label{4.23}
\psi_a(\vec r) = \left( \begin{array}{c} {}^{(a)}\!\varphi_+(\vec r) \\ {}^{(a)}\!\varphi_-(\vec r) \end{array}\right)
\end{equation}
and reformulating the mass eigenvalue equations in terms of the two-component Pauli spinors ${}^{(a)}\!\varphi_\pm(\vec r)$. In this way, the Dirac form (\ref{4.2}) of the first particles' eigenvalue equation is recast into the following Pauli form
\begin{multline}\label{4.24}
i \, \vec \sigma \bdot \vec \nabla {}^{(1)}\!\varphi_\pm(\vec r) + [ \exAo + \zAo + \dAo ] \cdot {}^{(1)}\!\varphi_\mp - \\
-[\vec A_2 + \vec A_3 ] \bdot \vec\sigma {}^{(1)}\!\varphi_\pm(\vec r) = \frac{\pm M_p - M_1}{\hbar}c\cdot {}^{(1)}\!\varphi_\mp(\vec r) \ ,
\end{multline}
where the usual Pauli matrices are denoted by $\vec \sigma = \{\sigma^j\}$. In a similar way, the two electronic eigenvalue equations (\ref{4.4a})-(\ref{4.4b}) appear in their Pauli form as
\begin{multline}\label{4.25}
i \, \vec \sigma \bdot \vec \nabla {}^{(2)}\!\varphi_\pm(\vec r) + [ \exAo + \eAo + \dAo ] \cdot {}^{(2)}\!\varphi_\mp + B_0 \cdot {}^{(3)}\!\varphi_\mp(\vec r) - \\
-[\vec A_1 + \vec A_3 ] \bdot \vec \sigma {}^{(2)}\!\varphi_\pm(\vec r) - \vec B \bdot \vec \sigma {}^{(3)}\!\varphi_\pm(\vec r) = -\frac{M_2 \pm M_e}{\hbar}c\cdot {}^{(2)}\!\varphi_\mp(\vec r) \ ,
\end{multline}
and
\begin{multline}\label{4.26}
i \, \vec \sigma \bdot \vec \nabla {}^{(3)}\!\varphi_\pm(\vec r) + [ \exAo + \eAo + \zAo ] \cdot {}^{(3)}\!\varphi_\mp + B_0^* \cdot {}^{(2)}\!\varphi_\mp(\vec r) - \\
-[\vec A_1 + \vec A_2 ] \bdot \vec \sigma {}^{(3)}\!\varphi_\pm(\vec r) - \vec B^* \bdot \vec \sigma {}^{(2)}\!\varphi_\pm(\vec r) = -\frac{M_3 \pm M_e}{\hbar}c\cdot {}^{(3)}\!\varphi_\mp(\vec r) \ .
\end{multline}
\indent But once the Pauli form of the eigenvalue equations is at hand, it is a simple matter to deduce thereof their non-relativistic approximation. Namely, one takes the equations for the upper Pauli components ${}^{(a)}\!\varphi_+(\vec r)$ in order to express approximately the lower components ${}^{(a)}\!\varphi_-(\vec r)$ in terms of the derivatives of the upper Pauli components, with simultaneous neglection of all the potential terms and putting also the mass eigenvalues $M_a$ ($a=1,2,3$) equal to the rest masses $M_p$ and $M_e$, resp. This yields for the first (positively charged) particle
\begin{equation}\label{4.27}
{}^{(1)}\!\varphi_-(\vec r) \approx \frac{i\hbar}{2\,M_p c} \, \vec \sigma \bdot \vec \nabla {}^{(1)}\!\varphi_+(\vec r) \ ,
\end{equation}
and similarly for the two electrons
\begin{subequations}\label{4.28}
\begin{align}
{}^{(2)}\!\varphi_-(\vec r) \approx -\frac{i\hbar}{2\,M_e c}\,\vec \sigma \bdot \vec \nabla {}^{(2)}\!\varphi_+(\vec r) \label{4.28a} \\
{}^{(3)}\!\varphi_-(\vec r) \approx -\frac{i\hbar}{2\,M_e c} \,\vec \sigma \bdot \vec \nabla {}^{(3)}\!\varphi_+(\vec r) \label{4.28b} \ .
\end{align}
\end{subequations}
Observe here again the change of sign for the positively charged particle (\ref{4.27}) relative to the electronic case (\ref{4.28a})-(\ref{4.28b}).

\indent The last step consists in substituting back these approximate lower components ${}^{(a)}\!\varphi_-(\vec r)$ into their corresponding eigenvalue equations (\ref{4.24})-(\ref{4.26}), in order to recover the well-known Schr\"odinger equations for the upper components ${}^{(a)}\!\varphi_+(\vec r)$; i.e. for the first particle
\begin{equation}\label{4.29}
-\frac{\hbar^2}{2\, M_p} \Delta {}^{(1)}\!\varphi_+(\vec r) + \hbar c \, [\exAo + \zAo + \dAo ] \cdot {}^{(1)}\!\varphi_+(\vec r) = E_{S(\bf 1)} \cdot {}^{(1)}\!\varphi_+(\vec r) \ ,
\end{equation}
and similarly for the two electrons:
\begin{subequations}\label{4.30}
\begin{multline}\label{4.30a}
-\frac{\hbar^2}{2\, M_e} \Delta {}^{(2)}\!\varphi_+(\vec r) - \hbar c \, [\exAo + \eAo + \dAo ] \cdot {}^{(2)}\!\varphi_+(\vec r) - \\ - \hbar c \, B_0 \cdot {}^{(3)}\!\varphi_+(\vec r) = E_{S(\bf 2)} \cdot {}^{(2)}\!\varphi_+(\vec r)
\end{multline}
\begin{multline}\label{4.30b}
-\frac{\hbar^2}{2\, M_e} \Delta {}^{(3)}\!\varphi_+(\vec r) - \hbar c \, [\exAo + \eAo + \zAo ] \cdot {}^{(3)}\!\varphi_+(\vec r) - \\ - \hbar c \, B^*_0 \cdot {}^{(2)}\!\varphi_+(\vec r) = E_{S(\bf 3)} \cdot {}^{(3)}\!\varphi_+(\vec r) \ .
\end{multline}
\end{subequations}
Indeed, these non-relativistic approximations of the exact mass eigenvalue equations do represent a very satisfying result because it is explicitly seen that all the electrostatic potentials $\exAo$, $\aAo$ couple to the positively charged particle (\ref{4.29}) with inverse sign in comparison to their coupling to the two electronic wave functions ${}^{(2,3)}\!\varphi_+(\vec r)$ (\ref{4.30a})-(\ref{4.30b}). Thus in this respect, the positivity and negativity of electric charge is correctly incorporated in the RST formalism. Furthermore, the Schr\"odinger energy eigenvalues $E_{S(\bf a)}$ do emerge correctly from the mass eigenvalues $M_a$:
\begin{subequations}\label{4.31}
\begin{align}
E_{S(\bf 1)} &= -(M_pc^2 + M_1 c^2) \label{4.31a} \\
E_{S(\bf 2)} &= M_2 c^2 - M_e c^2 \label{4.31b} \\
E_{S(\bf 3)} &= M_3 c^2 - M_e c^2 \label{4.31c} \ .
\end{align}
\end{subequations}
This outcome just meets with the fact that the non-relativistic Schr\"odinger energy eigenvalue $E_S$ of a bound particle must be always smaller than zero, but much larger than the negative rest mass energies ($-M_pc^2, -M_ec^2$).

\indent In a similar way, the consistency of the RST eigenvalue problem becomes evident also in connection with the non-relativistic approximation of the kinetic energies (\ref{4.22a})-(\ref{4.22c}). Indeed, one first rewrites these kinetic energies in terms of the Pauli spinors ${}^{(a)}\!\varphi_\pm(\vec r)$ (\ref{4.23}), e.g. for the first particle ($a=1$)
\begin{equation}\label{4.32}
T_{kin(\bf 1)} = - \frac{\hbar^2}{2\, M_p} \int d^3\vec r \; \left\{ {}^{(1)}\!\varphi_+^\dagger(\vec r) (\vec\sigma\bdot\vec\nabla) {}^{(1)}\!\varphi_-(\vec r) + {}^{(1)}\!\varphi^\dagger_-(\vec r) (\vec\sigma\bdot\vec\nabla) {}^{(1)}\!\varphi_+(\vec r) \right\} \ ,
\end{equation}
and then one introduces the non-relativistic approximation for the lower Pauli component (\ref{4.27}) in order to find
\begin{equation}\label{4.33}
T_{kin(\bf 1)} = - \frac{\hbar^2}{2\, M_p} \int d^3\vec r \; {}^{(1)}\!\varphi_+^\dagger(\vec r) \bdot \Delta {}^{(1)}\!\varphi_+(\vec r) \ .
\end{equation}
In the same way, the kinetic energies for the two electrons ($a=2,3$) are found to be of the following form
\begin{subequations}\label{4.34}
\begin{align}
T_{kin(\bf 2)}= -\frac{\hbar^2}{2\, M_e} \int d^3\vec r \; {}^{(2)}\!\varphi^\dagger_+(\vec r) \bdot \Delta  {}^{(2)}\!\varphi_+(\vec r) \label{4.34a} \\
T_{kin(\bf 3)}= -\frac{\hbar^2}{2\, M_e} \int d^3\vec r \; {}^{(3)}\!\varphi^\dagger_+(\vec r) \bdot \Delta  {}^{(3)}\!\varphi_+(\vec r) \label{4.34b} \ .
\end{align}
\end{subequations}

\indent Thus the non-relativistic approximation of the RST eigenvalue problem for an arbitrary collection of identical and non-identical particles is revealed to be in perfect agreement with the usual non-relativistic Schr\"odinger quantum mechanics, despite the fact that there intermediately emerge negative mass eigenvalues for the {\it relativistic} problem:
\begin{enumerate}
\item the relativistic mass eigenvalue equations (\ref{4.2})-(\ref{4.4b}) have the ordinary Schr\"odinger equations (\ref{4.29})-(\ref{4.30b}) as their non-relativistic approximations
\item the Schr\"odinger energy eigenvalues $E_{S(\bf a)}$ (\ref{4.31a})-(\ref{4.31c}) are correctly expressed by the relativistic mass eigenvalues $M_a$, be they positive or negative
\item the kinetic and potential energies of the particles adopt their usual non-relativistic form.
\end{enumerate}

\indent As the final step, it remains to be shown that the energy eigenvalues themselves are physically reasonable and do also agree with the general expectations, especially concerning their non-relativistic approximation. Since these energy eigenvalues emerge in RST as the values of the total energy functional $E_T$ upon the corresponding solutions of the eigenvalue problem, it becomes now necessary to inspect somewhat closer the latter functional.

\subsection{Energy Functional}
The energy of a bound state is in general (except for the one-particle case \cite{a27}) not identical to the sum of the mass eigenvalues ($\sum_a M_a c^2$) as they arise by solving the mass eigenvalue equations (\ref{4.2}) and (\ref{4.4a})-(\ref{4.4b}). The reason for this is that the two-particle interaction energies of both the electromagnetic and exchange type are contained already in anyone of the mass eigenvalues $M_a$ and therefore would be counted twice through forming simply the sum of mass eigenvalues. However this double-counting is corrected automatically if one defines the {\it total energy} $E_T$ of a bound state via the spatial integral of the total energy density ${}^{(T)}\!T_{00}(\vec{r})$ located in the RST field system:
\begin{equation}\label{4.35}
E_T = \int d^3\vec{r} \; {}^{(T)}\!T_{00}(\vec{r}) \ .
\end{equation}
Clearly, since the whole RST field system consists of a matter subsystem and a gauge field subsystem, the total energy density ${}^{(T)}\!T_{00}(\vec{r})$ must be the sum of a Dirac matter part (${}^{(D)}\!T_{00}$), internal gauge field part (${}^{(G)}\!T_{00}$) and the interaction contribution with respect to an external source (${}^{(es)}\!T_{00}$):
\begin{equation}\label{4.36}
{}^{(T)}\!T_{00}(\vec{r}) = {}^{(D)}\!T_{00}(\vec{r}) + {}^{(G)}\!T_{00}(\vec{r}) + {}^{(es)}\!T_{00}(\vec{r}) \ .
\end{equation}
Consequently, the total energy $E_T$ (\ref{4.35}) must also appear as a sum of three contributions
\begin{equation}\label{4.37}
E_T = E_D + E_G + E_{es} \ ,
\end{equation}
with the obvious identifications
\begin{subequations}\label{4.38}
\begin{align}
E_D &= \int d^3\vec{r} \; {}^{(D)}\!T_{00}(\vec{r}) \label{4.38a} \\
E_G &= \int d^3\vec{r} \; {}^{(G)}\!T_{00}(\vec{r}) \label{4.38b} \\
E_{es} &= \int d^3\vec{r} \; {}^{(es)}\!T_{00}(\vec{r}) \label{4.38c} \ .
\end{align}
\end{subequations}
The discussion of anyone of these three contributions yields now further confidence into the logical consistency of RST.

\subsubsection{External Interaction Energy $E_{es}$ $\textnormal(\ref{4.38c})$}
The point with the external interaction energy $E_{es}$ for different particles is here that the individual contributions to this quantity must be of opposite sign for oppositely charged particles (here: electrons and positrons / protons).

\indent In order to be convinced that this important requirement is actually satisfied by the present treatment of (non-)identical particles, one first recalls the energy-momentum density $\esTmunu$ of the external interactions \cite{a43}
\begin{equation}\label{4.39}
\esTmunu = - \frac{\hbar c}{\vpas} \Big \{ {}^{(ex)}\!F_{\mu \lambda} F_\nu^{\ \lambda} + {}^{(ex)}\!F_{\nu \lambda} F_\mu^{\ \lambda} - \frac{1}{2} g_{\mu\nu} {}^{(ex)}\!F_{\sigma\lambda} F^{\sigma\lambda} \Big\} \ .
\end{equation}
Splitting here up the external field strength $\exFmunu$ and the total field strength $\Fmunu$ (\ref{3.10}) into its electric ($\vec{E}$) and magnetic parts ($\vec{H}$) according to
\begin{subequations}\label{4.40}
\begin{align}
\vec{E} &= \{ E^j \} \doteqdot \{ F_{0j} \} \label{4.40a} \\
\vec{H} &= \{ H^j \} \doteqdot \{ \frac{1}{2} \, \epsilon^{jk}_{\ \ l} F_k^{\ l} \} \quad etc. \label{4.40b}
\end{align}
\end{subequations}
yields the following result for the energy density ${}^{(es)}\!T_{00}(\vec{r})$ of the external interactions
\begin{equation}\label{4.41}
{}^{(es)}\!T_{00}(\vec{r}) = \frac{\hbar c}{\vpas} \left\{ \vec{E}_{ex} \bdot \vec{E} + \vec{H}_{ex} \bdot \vec{H} \right\} \ .
\end{equation}
Thus the external energy density ${}^{(es)}\!T_{00}(\vec{r})$ is composed in a bilinear way of the external objects $\{ \vec{E}_{ex}, \vec{H}_{ex} \}$ and the total objects $\{ \vec{E}, \vec{H} \}$ of the system. The physical meaning of this result is that the system interacts as a whole (i.e. via its "total" quantities) with an external source.

\indent Next we restrict ourselves to a purely electric external source ($\vec{H}_{ex} \equiv 0$) and assume also that the external field $\vec{E}_{ex}$ is generated by an external electrostatic potential ${}^{(ex)}\!A_0$ in the usual (i.e Abelian) way
\begin{equation}\label{4.42}
\vec{E}_{ex}(\vec{r}) = -\vec{\nabla} {}^{(ex)}\!A_0(\vec{r}) \ ,
\end{equation}
just as is the case with the total potential $A_0(\vec{r})$ of the system:
\begin{equation}\label{4.43}
\vec{E}(\vec{r}) = -\vec{\nabla} A_0(\vec{r}) \ .
\end{equation}
Indeed the latter relation is nothing else than the time component of its Lorentzinvariant generalization (\ref{3.10}) which immediately is obtained by adding up the first three equations of the relationships (\ref{3.7a})-(\ref{3.7d}) between field strengths and potentials
\begin{equation}\label{4.44}
A_\mu \doteqdot \sum_{a=1}^3 A^a_{\ \mu} \ .
\end{equation}
But with the help of these arrangements, the external interaction energy $E_{es}$ (\ref{4.38c}) may be recast to the following form:
\begin{equation}\label{4.45}
E_{es} = - \hbar c \int d^3\vec{r} \; {}^{(ex)}\!A_0(\vec{r}) \cdot j_0(\vec{r}) \ ,
\end{equation}
where it is assumed that the total electrostatic potential $A_0(\vec r)$ is related to the total charge density $j_0(\vec r)$ via the usual Poisson equation
\begin{equation}\label{4.46}
\Delta A_0(\vec r) = \vpas \, j_0(\vec r) \ .
\end{equation}
Now recalling here the fact that the total charge density $j_0$ is composed of the three Dirac densities ${}^{(a)}\!k_0(\vec{r})$ ($a=1,2,3$) as specified by (c.f. (\ref{3.16}))
\begin{equation}\label{4.47}
j_0(\vec r) = -{}^{(1)}\!k_0(\vec r) + {}^{(2)}\!k_0(\vec r) + {}^{(3)}\!k_0(\vec r)
\end{equation}
immediately yields the result that $E_{es}$ (\ref{4.38c}) appears as the charge-weighted sum of the expected three contributions
\begin{equation}\label{4.48}
E_{es} = \sum_{a=1}^3 E_{es(\bf a)} \ ,
\end{equation}
with
\begin{subequations}\label{4.49}
\begin{align}
E_{es(\bf 1)} &=   \hbar c \int d^3\vec r \; {}^{(ex)}\!A_0(\vec{r}) \cdot {}^{(1)}\!k_0(\vec{r}) \label{4.49a} \\
E_{es(\bf 2)} &= - \hbar c \int d^3\vec r \; {}^{(ex)}\!A_0(\vec{r}) \cdot {}^{(2)}\!k_0(\vec{r}) \label{4.49b} \\
E_{es(\bf 3)} &= - \hbar c \int d^3\vec r \; {}^{(ex)}\!A_0(\vec{r}) \cdot {}^{(3)}\!k_0(\vec{r}) \label{4.49c} \ .
\end{align}
\end{subequations}
Thus the positively charged particle (\ref{4.49a}) contributes with the opposite sign in comparison to the two (negatively charged) electrons (\ref{4.49b})-(\ref{4.49c}); and this again validates the consistency of the present RST treatment of oppositely charged particles.

\subsubsection{Internal Gauge Field Energy $E_G$ $\textnormal{(\ref{4.38b})}$}
A characteristic feature of the RST {\it self-interactions} refers to the fact that they appear in connection with the gauge field modes $\Faomunu$ much more obvious than in connection with the matter fields $\psi_a$. The point here is that the particle self-interactions are encoded in the fibre metric $\Kaubu$ (see the discussion below equation (\ref{3.26})); and it is just this latter object which governs the way in which the individual gauge field modes $\Faomunu$ do cooperatively build up the internal energy-momentum density $\GTmunu$ \cite{a43}:
\begin{equation}\label{4.50}
\GTmunu = \frac{\hbar c}{\vpas} \Kaubu \{\Faomulu F^{\beta\ \lambda}_{\ \nu} - \frac{1}{4} \, g_{\mu\nu} \Faosulu F^{\beta\sigma\lambda}\} \ .
\end{equation}
The fibre metric $\Kaubu$ (\ref{3.23}) as the matrix of coupling constants consists essentially of three different elements, namely the {\it pair coupling constant} $K_p$
\begin{equation}\label{4.51}
K_p \doteqdot - e^u \cosh u \ ,
\end{equation}
the {\it self-coupling constant} $K_s$
\begin{equation}\label{4.52}
K_s \doteqdot e^u \sinh u \ ,
\end{equation}
and the {\it exchange coupling constant} $K_x$
\begin{equation}\label{4.53}
K_x \doteqdot - e^u ( \cosh u + 2 \sinh u ) \ .
\end{equation}
Therefore the energy density ${}^{(G)}\!T_{00}(\vec{r})$ as the time component of the energy-momentum density ${}^{(G)}\!T_{\mu\nu}$ must necessarily appear as a sum of the corresponding three subdensities:
\begin{align}\label{4.54}
{}^{(G)}\!T_{00}(\vec{r}) =&-K_p(u) \cdot \frac{\hbar c}{\vpas} \sum_{a\neq b} \{ \vec{E}_a \bdot \vec{E}_b + \vec{H}_a \bdot \vec{H}_b\}-\nonumber \\
&-K_s(u) \cdot \frac{\hbar c}{\vpas} \sum_{a=1}^3 \{ \vec{E}_a \bdot \vec{E}_a + \vec{H}_a \bdot \vec{H}_a \} + \nonumber \\
&+K_x(u) \cdot \frac{\hbar c}{\vpas} \{\vec{X}^* \bdot \vec{X} + \vec{Y}^* \bdot \vec{Y} \} \ .
\end{align}
\indent Here it should be rather evident that the first contribution ($\sim K_p(u)$) is the sum of mutual pair-interaction densities of the particles, the second contribution ($\sim K_s(u)$) is the particle self-energy density which appears here as the energy content of the real gauge field modes $\{\vec{E}_a,\vec{H}_a \}$ ($a=1,2,3$); and finally the third contribution ($\sim K_x(u)$) is the exchange energy density where the "electric" ($\vec{X}$) and "magnetic" ($\vec{Y}$) exchange field strengths are defined quite analogously to their electromagnetic counterparts $\vec{E}, \vec{H}$ (\ref{4.40a})-(\ref{4.40b}). Consequently the internal gauge field energy $E_G$ (\ref{4.38b}) must also consist of the corresponding three subenergies:
\begin{equation}\label{4.55}
E_G = \hat E_R + \tilde E_R -E_C \ .
\end{equation}
Here the pair-interaction energy $\hat E_R$ of the real gauge field modes is given by
\begin{equation}\label{4.56}
\hat E_R = -K_p(u) \cdot \frac{\hbar c}{\vpas} \sum_{a\neq b} \int d^3\vec{r} \; \{\vec{E}_a\bdot \vec{E}_b +\vec{H}_a \bdot \vec{H}_b \} \ ,
\end{equation}
similarly the self-energy  $\tilde E_R$ is obtained as
\begin{equation}\label{4.57}
\tilde E_R = -K_s(u) \cdot \frac{\hbar c}{\vpas} \sum_{a=1}^3 \int d^3\vec{r} \;\{\vec{E}_a\bdot \vec{E}_a +\vec{H}_a \bdot \vec{H}_a \} \ ,
\end{equation}
and finally the exchange energy $E_C$ appears as
\begin{equation}\label{4.58}
E_C = -K_x(u) \cdot \frac{\hbar c}{\vpas} \int d^3\vec{r} \; \{\vec{X}^* \bdot \vec{X} + \vec{Y}^* \bdot \vec{Y} \} \ .
\end{equation}
Obviously, anyone of these three contributions is itself the sum of its electric ($e,h$) and magnetic part ($m,g$) so that the internal gauge field energy $E_G$ (\ref{4.55}) actually is composed of six contributions:
\begin{equation}\label{4.59}
E_G = ( \hat E_R^{(e)} + \hat E_R^{(m)} ) + ( \tilde E_R^{(e)} + \tilde E_R^{(m)} ) - ( E_C^{(h)} + E_C^{(g)} ) \ .
\end{equation}
If the self-interaction parameter $u$ tends to zero ($u \rightarrow 0$), the self-energy $\tilde E_R$ (\ref{4.57}) tends to zero, too, because the self-interaction constant $K_s(u)$ (\ref{4.52}) vanishes in this case. What remains in this limit ($u \rightarrow 0$) is the mutual interaction energy $\hat E_R$ (\ref{4.56}) and the interaction energy $E_C$ (\ref{4.58}) where both coupling constants $K_p(u)$ and $K_x(u)$ tend to unity ($K_x(u) \rightarrow K_p(u) \Rightarrow -1$).

\subsubsection{Matter Energy $E_D$ $\textnormal{(\ref{4.38a})}$}
\indent Among the three energy contributions $E_{es}$, $E_G$, $E_D$ to the total energy $E_T$ (\ref{4.37}), the latter one ($E_D$) requires most of the attention. The crucial point here becomes readily obvious by a closer inspection of the right-hand sides of the mass eigenvalue equations (\ref{4.2}) and (\ref{4.4a})-(\ref{4.4b}). Indeed, whereas the two electronic equations (\ref{4.4a})-(\ref{4.4b}) are symmetric with respect to the interchange of the corresponding variables $\psi_2(\vec r) \leftrightarrow \psi_3(\vec r)$, $M_2 \leftrightarrow M_3$, the (positively charged) first particle (\ref{4.2}) appears to own a negative rest mass $M_p$! Therefore the question arises whether perhaps the emergence of such an apparently negative rest mass for the positively charged particle does spoil the physical consistency of the eigenvalue problem which would then imply that RST is not suited to describe a system of oppositely charged particles. However we shall readily demonstrate that just the opposite is true; namely the first particle contributes a positive term $E_{D(\bf 1)}$ to the matter energy $E_D$ (\ref{4.38a}). Moreover, the latter kind of energy is expected to consist of rest mass energy and of the kinetic energy of motion but nothing else.

\indent In order to verify this, observe first that the mass energy $E_D$ (\ref{4.38a}) of the three-particle system actually is the sum of three contributions due to any particle:
\begin{equation}\label{4.60}
E_D = E_{D(\bf 1)} + E_{D(\bf 2)} + E_{D(\bf 3)} \ .
\end{equation}
This circumstance becomes immediately obvious by simply inserting the covariant three-particle derivative (\ref{3.30}) in its component form (\ref{3.31a})-(\ref{3.31c}) into the energy-momentum density of matter $\DTmunu$ (\ref{2.50}) which then lets emerge the corresponding energy density ${}^{(D)}\!T_{00}(\vec{r})$ in the following form:
\begin{align}\label{4.61}
{}^{(D)}\!T_{00}(\vec{r}) =& \left[-M_1 c^2 - \hbar c \, ( {}^{(ex)}\!A_0 + {}^{(2)}\!A_0 + {}^{(3)}\!A_0 ) \right] \cdot {}^{(1)}\!k_0(\vec{r}) \nonumber \\
&+ \left[ M_2 c^2 + \hbar c \, ( {}^{(ex)}\!A_0 + {}^{(1)}\!A_0 + {}^{(3)}\!A_0 ) \right] \cdot {}^{(2)}\!k_0(\vec{r}) + \hbar c \, B_0 \cdot h_0(\vec{r}) \nonumber \\
&+ \left[ M_3 c^2 + \hbar c \, ( {}^{(ex)}\!A_0 + {}^{(1)}\!A_0 + {}^{(2)}\!A_0 ) \right] \cdot {}^{(3)}\!k_0(\vec{r}) + \hbar c \, B^*_0 \cdot h^*_0(\vec{r}) \ .
\end{align}
Therefore, since the mass equivalents $\hat z_1 \cdot M_I^{(e)} c^2$ (\ref{4.14}) etc. of the electrostatic and exchange interaction energies due to the potentials ${}^{(ex)}\!A_0$, ${}^{(a)}\!A_0$, $B_0$ are much smaller than the mass eigenvalues ($\sim M_ac^2$), the matter energy $E_D$ (\ref{4.60}) is dominated by just those mass eigenvalues of the particles:
\begin{subequations}\label{4.62}
\begin{align}
E_{D(\bf 1)} &= - \hat z_1 \cdot (M_1 + M_I^{(e)} ) c^2 - E_{es(\bf 1)} = - \hat z_1 \cdot M_1c^2 + \dots \label{4.62a} \\
E_{D(\bf 2)} &= - \hat z_2 \cdot (-M_2 +M_{I\!I}^{(e)} + M_h ) c^2 - E_{es(\bf 2)} = \hat z_2 \cdot M_2c^2 + \dots \label{4.62b} \\
E_{D(\bf 3)} &= - \hat z_3 \cdot (-M_3 +M_{I\!I\!I}^{(e)} + M^*_h ) c^2 - E_{es(\bf 3)} = \hat z_3 \cdot M_3c^2 + \dots \label{4.62c} \ ,
\end{align}
\end{subequations}
where the normalization paramters $\hat z_a$ ($a=1,2,3$) are given by equation (\ref{4.7}). Thus it becomes immediately obvious from the present conclusion (\ref{4.62a}) that the first mass eigenvalue $M_1$ {\it must} be negative in order that its contribution ($-M_1 c^2$) to the total matter energy $E_D$ be {\it positive}!

\subsection{Poisson Equations}
\indent The mass eigenvalue equations (\ref{4.2}) and (\ref{4.4a})-(\ref{4.4}) do not yet form a closed system; and therefore they must be complemented by the field equations for the gauge potentials $\{ \aAo, \vec A_a; B_0, \vec B \}$ in order to close the eigenvalue system. The desired Poisson equations for the gauge potentials are to be deduced from the Maxwell equations (\ref{3.9a})-(\ref{3.9d}) by substituting therein the field strengths $\Famunu, \Gmunu$ due to the corresponding potentials as specified by equations (\ref{3.7a})-(\ref{3.7d}). Since the potentials do enter the mass eigenvalue system in three-vector form, it is convenient to first display also the system (\ref{3.7a})-(\ref{3.7d}) in three-vector notation. Introducing here the three-vectors $\vec E_a, \vec H_a, \vec X, \vec Y$ as being defined by equations (\ref{4.40a})-(\ref{4.40b}) yields then the desired relationship between the electric field strengths and potentials in the following form:
\begin{subequations}\label{4.64}
\begin{align}
\vec E_1(\vec r) &= - \vec \nabla \eAo(\vec r) \label{4.64a} \\
\vec E_2(\vec r) &= - \vec \nabla \zAo(\vec r) - i \, \big[B_0(\vec r) \vec B^*(\vec r) - B_0^*(\vec r) \vec B(\vec r) \big] \label{4.64b} \\
\vec E_3(\vec r) &= - \vec \nabla \dAo(\vec r) + i \, \big[B_0(\vec r) \vec B^*(\vec r) - B_0^*(\vec r) \vec B(\vec r) \big] \label{4.64c} \\
\vec X(\vec r)   &= - \vec \nabla B_0(\vec r) + i \, B_0(\vec r) \, \big[ \vec A_2(\vec r) - \vec A_3(\vec r) \big] + i \, \Delta_0(\vec r) \vec B(\vec r) \label{4.64d}
\end{align}
\end{subequations}
where the electrostatic potential difference $\Delta_0(\vec r)$ is defined through
\begin{align}\label{4.65}
\Delta_0(r) \doteqdot &\frac{1}{a_M} - \left [ {}^{(2)}\!A_0(r) - {}^{(3)}\!A_0(r) \right ] \\
\Big( a_M &\doteqdot \frac{\hbar}{(M_2 -M_3)c} \Big) \ . \nonumber
\end{align}
The corresponding relationships for the magnetic objects read
\begin{subequations}\label{4.66}
\begin{align}
\vec H_1(\vec r) &= \vec \nabla \times \vec A_1(\vec r) \label{4.66a} \\
\vec H_2(\vec r) &= \vec \nabla \times \vec A_2(\vec r) - i \, \vec B(\vec r) \times \vec B^*(\vec r) \label{4.66b} \\
\vec H_3(\vec r) &= \vec \nabla \times \vec A_3(\vec r) + i \, \vec B(\vec r) \times \vec B^*(\vec r) \label{4.66c} \ ,
\end{align}
\end{subequations}
and similarly for the "magnetic" exchange object $\vec Y(\vec r)$
\begin{equation}\label{4.67}
\vec Y(\vec r) = \vec \nabla \times \vec B(\vec r) - i \, \big [\vec A_2(\vec r) - \vec A_3(\vec r) \big ] \times \vec B(\vec r) \ .
\end{equation}

With these preparations, the desired Poisson equations may be obtained now by simply substituting the present three-vector form of the curvature components into the corresponding three-vector form of the Maxwell equations (\ref{3.9a})-(\ref{3.9d}). The electric part of these equations appears in the following form
\begin{subequations}\label{4.68}
\begin{align}
&\vec \nabla \bdot \vec E_1 = \vpas \, {}^{(1)}\!j_0 \label{4.68a} \\
&\vec \nabla \bdot \vec E_2 + i \, \big[ \vec B^* \bdot \vec X - \vec B \bdot \vec X^* \big ] = \vpas\,{}^{(2)}\!j_0 \label{4.68b} \\
&\vec \nabla \bdot \vec E_3 - i \, \big[ \vec B^* \bdot \vec X - \vec B \bdot \vec X^* \big ]= \vpas\, {}^{(3)}\!j_0 \label{4.68c} \\
&\vec \nabla \bdot \vec X - i \, \big [\vec A_2 - \vec A_3 \big] \bdot \vec X +i\, \vec B\bdot \big[\vec E_2-\vec E_3\big] = \vpas \, g_0 \label{4.68d}
\end{align}
\end{subequations}
where the Maxwellian charge densities ${}^{(a)}\!j_0(\vec r)$ and $g_0$ are the time components of the four-currents $j^\alpha_{\ \mu}$ (\ref{3.25a})-(\ref{3.25e}). As a consistency check, one adds up here the first three equations (\ref{4.68a})-(\ref{4.68c}) in order to find the source equation for the total electric field $\vec E (\doteqdot \vec E_1 + \vec E_2 + \vec E_3 )$ as
\begin{equation}\label{4.69}
\vec \nabla \bdot \vec E = - \vpas \, j_0(\vec r)
\end{equation}
where the total charge density $j_0(\vec r)$ is the time component of the total four-current $j_\mu$ (\ref{3.16}). Clearly, the source equation (\ref{4.69}) is nothing else than the electric part of the total Maxwell equation (\ref{3.11}).

\indent The magnetic counterparts of the electric source equations are to be deduced from the Lorentz covariant form (\ref{3.9a})-(\ref{3.9d}) in a quite analogous way and are then found to look as follows
\begin{subequations}\label{4.70}
\begin{align}
&\vec \nabla \times \vec H_1 = \vpas \, \vec j_1 \label{4.70a} \\
&\vec \nabla \times \vec H_2 + i \, \big[ \vec B^* \times \vec Y - \vec B \times \vec Y^* \big] - i \, \big[ B_0 \vec X^* - B^*_0 \vec X \big] = \vpas \, \vec j_2 \label{4.70b} \\
&\vec \nabla \times \vec H_3 - i \, \big[ \vec B^* \times \vec Y - \vec B \times \vec Y^* \big] + i \, \big[ B_0 \vec X^* - B^*_0 \vec X \big] = \vpas \, \vec j_3 \label{4.70c} \\
&\vec \nabla \times \vec Y + \frac{i}{a_M} \vec X - i \, \big[ \vec A_2 - \vec A_3 \big] \times \vec Y + i \, B_0 \big[ \vec E_2 - \vec E_3 \big] -i \, \big[ \vec H_2 - \vec H_3 \big] \times \vec B = \vpas \, \vec g \ . \label{4.70d}
\end{align}
\end{subequations}
Clearly, the sum of the first three equations (\ref{4.70a})-(\ref{4.70c}) is again the magnetic part of the total Maxwell equation (\ref{3.11}) and reads
\begin{equation}\label{4.71}
\vec \nabla \times \vec H = - \vpas \, \vec j
\end{equation}
with the total three-current $\vec j$ being the spatial part of $j_\mu$ (\ref{3.16}).

\indent The present three-vector form of the {\it non-Abelian} Maxwell equations demonstrates, in an even more impressive way than the original Lorentz covariant form (\ref{3.9a})-(\ref{3.9d}), the amount of intricacy which is brought in by the {\it non-Abelian} character of the theory. On the other hand, the non-linearities due to the non-Abelian structure have only a small influence upon the observable quantities of the theory (i.e. energy levels etc.). In fact, the non-linear effects may be of the order of magnitude of the relativistic and self-interaction effects, so that one may neglect the non-linear terms together with the self-interactions ($u\rightarrow 0$) if one whishes to pass over to the non-relativistic limit of the theory. In this sense, one linearizes the non-Abelian gauge field equations and thus finds the following {\it linear} Poisson equations of the electric type
\begin{subequations}\label{4.72}
\begin{align}
&\Delta \eAo = - \vpas \, \eko \label{4.72a} \\
&\Delta \zAo = \vpas \, \zko \label{4.72b} \\
&\Delta \dAo = \vpas \, \dko \label{4.72c} \\
&\Delta B_0  = - \vpas \, h_0^* \label{4.72d} \ ,
\end{align}
\end{subequations}
and similarly of the magnetic type:
\begin{subequations}\label{4.73}
\begin{align}
&\Delta \vec A_1 = - \vpas \, \vec k_1 \label{4.73a} \\
&\Delta \vec A_2 = \vpas \, \vec k_2 \label{4.73b} \\
&\Delta \vec A_3 = \vpas \, \vec k_3 \label{4.73c} \\
&\Delta \vec B  = - \vpas \, \vec h^* \label{4.73d} \ .
\end{align}
\end{subequations}

The sources of the gauge potentials are emerging here in form of the Dirac currents $k_{a\mu}$ (\ref{3.13a})-(\ref{3.14b}) because the Maxwell currents $j^\alpha_{\ \mu}$ (\ref{3.25a})-(\ref{3.25e}) reduce to the Dirac currents $k_{a\mu}$ for vanishing self-interactions ($u\rightarrow0$). For the localized atomic configurations to be considered here, one assumes that the wave functions $\psi_a(\vec r)$ vanish at spatial infinity (i.e. $\vert \vec r \vert \rightarrow \infty$) which then also holds for the gauge potentials. Therefore, one may adopt here the standard solutions of the linear Poisson equations, such as e.g. for the exchange potential $B_0(\vec r)$ (\ref{4.72d})
\begin{equation}\label{4.74}
B_0(\vec r) = \alpha_{\rm S} \int d^3\vec r\,' \; \frac{h_0^*(\vec r\,')}{\vert \vec r-\vec r\,' \vert} \ , \textnormal{ etc.}
\end{equation}

\indent The latter result is well-suited in order to estimate the magnitude of the exchange energy $E_C$ (\ref{4.58}). Observing here the fact that the exchange energies of the "magnetic type" $E_C^{(g)}$ are mostly much smaller than their "electric" counterparts $E_C^{(h)}$ (putting $u\rightarrow 0$):
\begin{equation}\label{4.75}
E_C^{(g)} \doteqdot \frac{\hbar c}{\vpas} \int d^3\vec r \; \vec Y^*(\vec r) \bdot \vec Y(\vec r) \ll E_C^{(h)} \doteqdot \frac{\hbar c}{\vpas} \int d^3\vec r \; \vec X^*(\vec r) \bdot \vec X(\vec r) \ ,
\end{equation}
we can approximately identify the total exchange energy $E_C$ (\ref{4.58}) with its "electric" part $E_C^{(h)}$, i.e.
\begin{equation}\label{4.76}
E_C \sim E_C^{(h)} = e^2 \int \!\!\!\! \int d^3\vec r \, d^3\vec r\,' \; \frac{h_0(\vec r)\cdot h_0^*(\vec r\,')}{\vert \vec r - \vec r\,' \vert} \ .
\end{equation}
Indeed this approximative result is easily deduced from the exact relation (\ref{4.75}) by resorting to the linearized approximation for the "electric" exchange field strength $\vec X(\vec r)$ (\ref{4.64d}), i.e.
\begin{equation}\label{4.77}
\vec X(\vec r) \approx -\vec \nabla B_0(\vec r) \ ,
\end{equation}
and by using also the linearized version (\ref{4.72d}) of the exchange Poisson equation. Furthermore, retaining for the non-relativistic approximation only the upper Pauli components ${}^{(a)}\!\varphi_+(\vec r)$ ($a=2,3$) of the Dirac spinors $\Psi_a(\vec r)$ (\ref{4.23}), the exchange density $h_0(\vec r)$ as the time component of the exchange current  $h_\mu$ (\ref{3.14a})-(\ref{3.14b}) is found to read in terms of the relevant Pauli components ${}^{(a)}\!\varphi_+(\vec r)$ as follows
\begin{equation}\label{4.78}
h_0(\vec r) \Rightarrow {}^{(2)}\!\varphi^\dagger_+(\vec r) \bdot {}^{(3)}\!\varphi_+(\vec r) \ .
\end{equation}
Consequently, the non-relativistic exchange energy $E_C^{(h)}$ (\ref{4.76}) is finally given by
\begin{equation}\label{4.79}
E_C^{(h)} \approx e^2 \int\!\!\!\!\int d^3\vec r \, d^3\vec r\,' \; \frac{ \Big( {}^{(2)}\!\varphi^\dagger_+(\vec r) \bdot {}^{(3)}\!\varphi_+(\vec r) \Big) \cdot \Big( {}^{(3)}\!\varphi^\dagger_+(\vec r\,') \bdot {}^{(2)}\!\varphi_+(\vec r\,') \Big)}{\vert \vec r - \vec r\,' \vert} \ .
\end{equation}

\subsection{Exchange Doublets}
The approximative result (\ref{4.79}) for the exchange energy of identical particles is known in the standard theory as "exchange integral" \cite{b31}, where the Pauli spinors may be replaced by the Schr\"odinger wave functions. But this exchange integral determines the exchange energy in both approaches in a different way:
\begin{enumerate}
\item In the standard theory, the Schr\"odinger energy eigenvalue $E_S$ of the two bound electrons ($a=2,3$) would be given by \cite{b31}
\begin{equation}\label{4.80}
E_S^{(\pm)} = E_S^{(0)} + E_R^{(e)} \pm E_C^{(h)} \ .
\end{equation}
Here $E_S^{(0)}$ is the energy of both non-interacting electrons, $E_C^{(h)}$ is the standard version of the RST exchange energy (\ref{4.79}), and $E_R^{(e)}$ is the mutual electrostatic interaction energy as the standard version of the RST energy $\hat E_R^{(e)}$ (\ref{4.56}):
\begin{equation}\label{4.81}
E_R^{(e)} = \frac{\hbar c}{\vpas} \int d^3\vec r \; \vec E_2 \bdot \vec E_3 = e^2 \int \!\!\!\! \int d^3\vec r \, d^3\vec r\,' \; \frac{{}^{(2)}\!k_0(\vec r\,') {}^{(3)}\!k_0(\vec r)}{\vert \vec r - \vec r\,' \vert} \ ,
\end{equation}
with the Dirac densities ${}^{(a)}\!k_0(\vec r)$ (\ref{4.9a}) being approximated by the upper Pauli components ${}^{(a)}\!\varphi_+(\vec r)$ (\ref{4.23}) as
\begin{align}\label{4.82}
{}^{(a)}k_0(\vec r) \Rightarrow& {}^{(a)}\!\varphi^\dagger_+(\vec r) \bdot {}^{(a)}\!\varphi_+(\vec r)\\
&(a=2,3) \ . \nonumber
\end{align}
Clearly, the {\it Coulomb integral} (\ref{4.81}) comes about in RST by resorting to the preceding linear approximations (\ref{4.68b})-(\ref{4.68c}), (\ref{4.70b})-(\ref{4.70c}) which entail the linear Poisson equations (\ref{4.72b})-(\ref{4.72c}). From the standard result (\ref{4.80}) one concludes that the spacing $\Delta E_S$ of the exchange doublet is
\begin{equation}\label{4.83}
\Delta E_S = E_S^{(+)} - E_S^{(-)} = 2 \, E_C^{(h)}
\end{equation}
where the upper (lower) sign in (\ref{4.80}) refers to antiparallel/parallel spins.
\item However, in RST, the lowest order approximation of the total energy $E_T$ is given by
\begin{equation}\label{4.84}
E_{RST} = E_{RST}^{(0)} + E_R^{(e)} - E_C^{(h)} \ .
\end{equation}
Thus in contrast to the standard result $E_S^{(\pm)}$ (\ref{4.80}), the RST exchange contribution is always negative for both parallel (singlet state) and antiparallel spins (triplet states). This is in qualitative agreement with the experimental data though being in contradiction to the standard result (\ref{4.80}). The spacing of the RST exchange doublet appears to be only half the value $\Delta E_S$ (\ref{4.83}) of the standard approach:
\begin{align}\label{4.85}
\Delta E_{RST} &= - \Big[ E_C^{(h)}\big\vert_{singlet} - E_C^{(h)}\big\vert_{triplet} \Big ] \\
	       &\Rightarrow E_C^{(h)} \big\vert_{triplet} \nonumber
\end{align}
because the exchange energy $E_C^{(h)}$ vanishes for the singlet state due to the orthogonality of the Pauli spinors ${}^{(2)}\!\varphi_+, {}^{(3)}\!\varphi_+$ in (\ref{4.79}) for antiparallel spins (see the discussion of this effect in ref \cite{a35}.
\end{enumerate}

\indent Thus the lowest order of approximations of both the standard approach and RST are afflicted with certain deficiencies and it remains to be shown that the spacing deficiency of RST (\ref{4.85}) disappears in the higher-order approximations in a similar way as the lowering problem (upper sign in (\ref{4.80}) disappears in the conventional theory by resorting to the multi-configuration Dirac-Fock method (MCDF) \cite{b15}.

\bibliographystyle{prsty}
\bibliography{bibdb_2}

\end{document}